\def\funp{{I\!\!P}}

\def\gapprox{\lower .7ex\hbox{$\;\stackrel{\textstyle >}{\sim}\;$}}
\def\lapprox{\lower .7ex\hbox{$\;\stackrel{\textstyle <}{\sim}\;$}}
\def\bfr{{\bf r}}

\newcommand{\be}{\begin{equation}}
\newcommand{\ee}{\end{equation}}
\newcommand{\bea}{\begin{eqnarray}}
\newcommand{\beqn}{\begin{eqnarray}}
\newcommand{\eea}{\end{eqnarray}}
\newcommand{\eeqn}{\end{eqnarray}}
\newcommand{\fP}{I\!\!P}\newcommand{\rf}{{\bf r}}
\newcommand{\fr}{{\bf r}}
\newcommand{\kf}{{\bf k}}
\newcommand{\fk}{{\bf k}}
\newcommand{\fl}{{\bf l}}
\newcommand{\lf}{{\bf l}}

\documentstyle[12pt,epsfig]{article}
\newlength{\dinwidth}
\newlength{\dinmargin}
\begin{document}

\titlepage
\begin{flushright}
DTP/99/20\\
March 1999
\end{flushright}

\vspace*{1in}
\begin{center}
{\Large \bf Saturation in Diffractive Deep Inelastic Scattering}
\vspace*{0.5in}\\
K. \ Golec-Biernat$^{2,1}$
and M. W\"usthoff$^{1}$ \\
\vspace*{0.5cm}
$^1${\it Department of Physics, University of Durham, Durham DH1 3LE, UK} \\
\vspace*{0.2cm}
$^2${\it H.\ Niewodniczanski 
Institute of Nuclear Physics, 
Department of Theoretical Physics, Radzikowskiego 152, Krakow, Poland}
\end{center}
\vspace*{2cm}
%\centerline{(\today)}

\vskip1cm 
\begin{abstract}
We successfully describe the HERA-data on diffractive deep inelastic scattering
using a saturation model which has been applied in our earlier analysis
of the inclusive $ep$-scattering data.
No further parameters are needed.
Saturation already turned out to be essential in describing the transition 
from large to small values of $Q^2$ in inclusive scattering. 
It is even more important for diffractive processes and naturally leads to a constant ratio 
of the diffractive versus inclusive cross sections. We present an extensive 
discussion of our results as well as detailed comparison with data.
\end{abstract}

%%%%%%%%%%%%%%%%%%%%%%%%%%%%%%%%%%%%%%%%%%%%%%%%%%%%%%%%%%%%%%%%%%%%%%%%%%%%%%%
\newpage
\section{Introduction}
In a recent analysis \cite{GBW} we introduced a model  which 
provides a description of the transition between large and low $Q^2$ 
in inclusive lepton--proton deep inelastic scattering at low $x$. 
The idea behind our model is a phenomenon which we call
a combined saturation at low $Q^2$ and low $x$.
In this kinematical region the size of the virtual probe 
is of the order of the mean transverse distance between partons in the proton.
The cross section for the interaction between the probe and the partons
becomes large and multiple scattering has to be taken into account.
These effects lead to saturation of the total cross section.
%Low-$x$ saturation occurs in the semi-hard region where the 
%parton density becomes very high and recombination effects limit
%the further growth of the density. The consequence is a reduction of the 
%growth of the cross section at small $x$. 
We found that saturation occurs
at low but still perturbative values of $Q^2$ ($\sim 1-2~GeV^2$ for 
$x=10^{-4}$). We therefore believe that saturation should be described
by means of perturbative QCD  
(see also for example Ref.~\cite{GLR,Mueller1,LA,Kovner}).
%It is necessary, however, to go beyond
%linear evolution schemes and take into account non-linear effects.

The QCD-framework we use allows us
to describe not only  inclusive but diffractive processes as well.
A general feature of diffraction is its strong 
sensitivity towards the infrared regime even for large $Q^2$. 
The fact that diffraction has a strong
soft component has already been noticed earlier, leading to the assumption
that the Pomeron in diffraction ought to be soft. The idea of saturation,
however, emphasizes the transition from hard to soft physics. 
As mentioned earlier saturation
effects become already viable at rather hard scales and strongly 
suppress soft contributions in diffractive processes \cite{Mue}. 
This mechanism leads to an effective
enhancement of hard contributions and hence to an effective
Pomeron intercept which lies above the original soft value.

The important conclusion of this paper is that the concept of saturation
leads to a good 
description of the diffractive data. Our approach has the important property
that the inclusive and diffractive cross section have the same power-behavior
in $x$. We have obtained these results 
without the use of any additional fitting parameters, i.e.
we solely take the model as determined in Ref.~\cite{GBW} from the
analysis of inclusive processes.
The diffractive slope which we use is taken from the measurement at HERA. 

The plan of our paper is as follows. In Section 2 and 3
we recapitulate the results found in Ref.~\cite{GBW} 
and discuss qualitatively the basic features
of saturation for both inclusive and diffractive processes.
In Section 4 we provide the detailed formulae for our numerical
analysis and discuss the comparison with the data  from 
H1 \cite{H1dif} and ZEUS \cite{ZEUSdif} in Section 5.
In Section 6 we present the impact parameter version 
of our results and finish with  concluding remarks in Section 7.
The Appendix was added to enclose some details on the derivation
of the cross section formulae for diffractive scattering. In particular
it contains the computation
of the relevant set of Feynman diagrams
which contribute to the quark-antiquark-gluon final state.

%%%%%%%%%%%%%%%%%%%%%%%%%%%%%%%%%%%%%%%%%%%%%%%%%%%%%%%%%%%%%%%%%%%%%
\section{Saturation in inclusive processes}

Our starting point in Ref.~\cite{GBW} was the well established
physical picture of small-$x$ interactions 
in which the photon with virtuality $Q^2$, emitted by a lepton, 
dissociates into a quark-antiquark pair far 
upstream of the nucleon (in the nucleon rest frame). The dissociation
is then followed by
the scattering of the  quark-antiquark pair on the nucleon. 
In this picture the relative transverse separation ${\bf r}$ of the
$q\bar{q}$ pair and the longitudinal photon momentum fraction $\alpha$
of the quark ($1-\alpha$ for the antiquark) 
are good degrees of freedom. In these variables 
the $\gamma^*p$ cross sections 
have the following factorized form \cite{NIK1,JEFF}
\be
\label{eq:1}
\sigma_{T,L}(x,Q^2)\:=\:\int \!d\,^2{\bf r}\! \int_0^1 \!d\alpha \:  
\vert \Psi_{T,L}\,(\alpha,\bfr) \vert ^2 \: \hat{\sigma}\,(x,r^2)\;,
\\
\ee
where $\Psi_{T,L}$ is the squared photon wave function 
for the transverse $(T)$ and longitudinally polarized  $(L)$ photons,
given by
\be
\label{eq:2}
\vert \Psi_{T,L}\,(\alpha,\bfr) \vert^2  \,=\,  
\frac{6\, \alpha_{em}}{4\,\pi^2 } \, \sum_{f}\,e_f^2 \, \cases { 
[\,\alpha^2+(1-\alpha)^2\,]\:\overline{Q}^2\,K_1^2\,(\overline{Q}\, r) 
\,+\, m_f^2\: K_0^2\,(\overline{Q}\, r)\,
\cr\cr
4\,Q^2\,\alpha^2\,(1-\alpha)^2\,K_0^2\,(\overline{Q}\, r)\;.
}
\ee
In the above formulae $K_0$ and $K_1$ are Mc Donald-Bessel  functions
and
\be
\label{eq:4}
\overline{Q}^2 \:=\:  \alpha\,(1-\alpha)\,Q^2\,+\,m_f^2\;.
\ee
The dynamics of saturation is embedded 
in the the effective dipole cross section
$\hat{\sigma}(x,r)$ which describes the interaction of the $q\bar{q}$ pair with
a nucleon:
\bea \label{sigmahat}
\hat{\sigma}(x,r^2)\,=\,\sigma_0\,\left \{1\,-\,
\exp\left(-\frac{r^2}{4 R_0^2(x)}\right) 
\right\}\;,
\eea
where the $x$-dependent radius $R_0$ is given by
\bea\label{R0}
R_0(x)&=&\frac{1}{GeV}\;\left(\frac{x}{x_0}\right)^{\lambda/2} \;.
\eea
The normalization $\sigma_0$ and 
the parameters $x_0$ and $\lambda>0$ of $R_0(x)$ have been determined
by a fit to all inclusive data on  $F_2$ with $x < 0.01$~ \cite{GBW}.
(the detailed values of these parameters are quoted in Section~5). 
%\input fig1
%%%%%%%%%%%%%%%%%%%%%%%%%%%
\begin{figure}[t] 
  \vspace*{-1cm}
     \centerline{
         \epsfig{figure=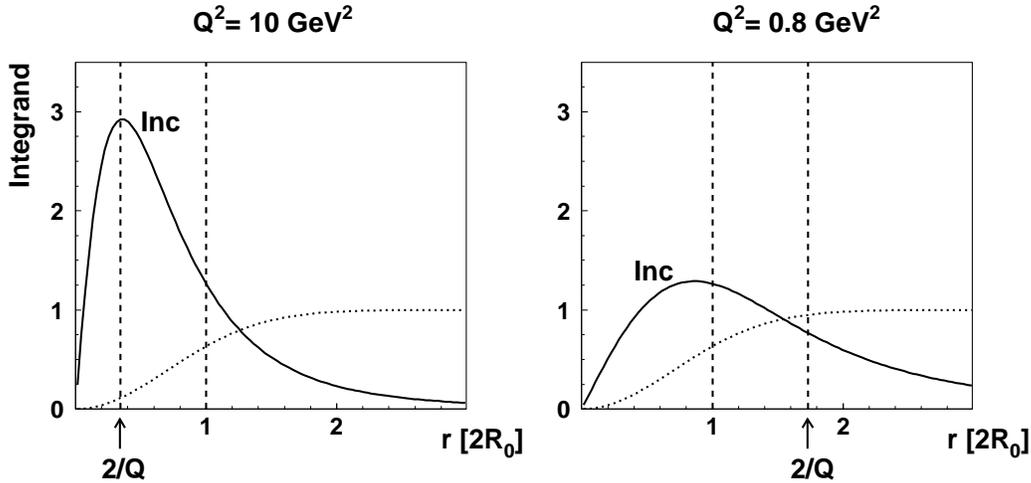,width=15cm}
           }
\vspace*{0.0cm}
\caption{\it The integrand 
of the inclusive cross section $\sigma_T$ in (\ref{eq:1}) (solid lines)
after the integration over $\alpha$ and
the azimuthal angle, plotted for two values of $Q^2$. 
The dotted lines
show the dipole cross section (\ref{sigmahat}). The dashed vertical lines 
correspond to the characteristic scales 
{\mbox{\rm $r=2 R_0$}} and {\mbox{\rm $r=2/Q$}}. The values for $2 R_0$ at
a fixed energy $W=245~\mbox{\rm GeV}$ are: 
$0.36~\mbox{\rm fm}$ for $Q^2=10~\mbox{\rm GeV}^2$
and $0.25~\mbox{\rm fm}$ for $Q^2=0.8~\mbox{\rm GeV}^2$. 
\label{rint1}}
\end{figure}
%%%%%%%%%%%%%%%%%%%%%%%%%%%

Saturation in the dipole cross section (\ref{sigmahat})
sets in when  $r \sim 2R_0$,  allowing a good description
of the data at low $Q^2$ when $1/Q > R_0$ (see right plot in Fig.~\ref{rint1}).
The detailed analysis of Eqs.~(\ref{eq:1})-(\ref{sigmahat}), presented 
in Ref.~\cite{GBW}, gives for small $Q^2$
\be
\label{eq:5}
\sigma_{T}\,\sim\,\sigma_0\,.
\ee
For large $Q^2$ the dominant contribution comes
from small dipole configurations with  $r \sim 2/Q \ll R_0$
(see left plot in Fig.~\ref{rint1}). In this case
we have the usual situation of color transparency,
$\hat{\sigma} \sim r^2$, which gives the scaling behavior of the
$\gamma^*p$ cross sections
\be
\label{eq:6}
\sigma_T \sim {1}/{Q^2}\,.
\ee
More precise analysis leads to logarithmic in $Q^2$ modifications
of the above estimations. 
%Logarithmic corrections due to QCD evolution have been neglected in order
%to keep the model simple. 
To summarize, the saturation model naturally interpolates
between the two different regimes of $\sigma_T$ described
by Eqs.~(\ref{eq:5}) and (\ref{eq:6}).

Saturation is characterized by a `critical line'\footnote{
There is no phase transition or critical behavior present in our approach.}
in the $(x,Q^2)$-plane given by the equation $Q^2=1/R_0^2(x)$ \cite{GBW,GLR,LA}. 
It is important to note that at very small $x$ saturation effects
become relevant at fairly high scales ($Q^2 \sim 1-2~GeV^2$ for HERA energies
\cite{GBW}) where one believes that perturbative QCD is valid.
The critical line divides the phase space into two regions, the scaling 
region in which relation (\ref{eq:6}) is valid
and the saturation region with the behavior given  by Eq.~(\ref{eq:5}).

The physical picture behind saturation is based on interpretation
of the $x$-dependent radius $R_0(x)$ as the mean separation of partons
in the proton.
We see from  (\ref{R0}) that when $x$ decreases so does the mean seperation.
Thus at low $x$ the distribution of partons 
in the proton is no longer dilute when probed
by a virtual photon of fixed resolution ($\sim1/Q$) and saturation sets in.
This happens 
when the resolution of the probe equals to the mean separation, $1/Q=R_0(x)$, 
which condition defines the "critical line".
As a result the dipol cross section becomes large and mulitple interactions
become important. In other words, at low $x$  proton appears
to be black. The important result of our inclusive analysis 
is that blackening occurs 
already at rather short distances well below where 'soft dynamics'
is supposed to set in, justifying the use of perturbative QCD.

In the scaling region of large $Q^2$ the growth of the inclusive 
cross section is driven by the increase in the number of partons 
since the gluon density $G(x)$ is proportional to $1/R_0^2(x)$ (see
Section 4 for detailes of this relation). This grows is eventually tamed
in our model by the mechanism of saturation.

%%%%%%%%%%%%%%%%%%%%%%%%%%%%%%%%%%%%%%%%%%%%%%%%%%%%%%%%%%%%%%%%%%%%
\section{Saturation in diffractive deep inelastic scattering}

Inclusive $\gamma^*p$ cross section at large $Q^2$ is dominated by the scaling
region. Diffractive scattering on the other hand
is essentially determined by the saturation region. In this case the dependence
on $x$ is controlled by the available phase space in the transverse momentum.
This phase space grows proportional to $1/R_0^2(x)$ 
and leads to the same power behavior
in $x$ as was found for the inclusive cross 
section. It also means that the average transverse momentum
of the diffractive final state will increase when $x$ decreases.
The process becomes 'harder' when  $x$  becomes smaller. 
Crucial for this picture to work is the scale invariance which in our
approach is maintained by the lack of any additional
cutoff on the transverse momenta of the final state. We will discuss this
conclusion in more details below.

In order to demonstrate the main features of saturation in diffraction 
we will confine our
discussion in this section to the elastic scattering of the $q\bar{q}$-pair
as shown in Fig.~\ref{diag}a. Elastic $q\bar{q}$-scattering 
dominates diffractive $\gamma^*p$ scattering 
for not to large values of the diffractive mass $M$.
At large $M$, however, the emission of a gluon
as depicted in Fig.~\ref{diag}b becomes the dominant contribution.  
The cross section for elastic $q\bar{q}$-scattering takes on 
the following form
\be\label{eq:diff1}
\left. \frac{d\sigma_{T,L}^D}{d t}\right|_{t=0}\;=\;\frac{1}{16\pi}\;
\int \!d\,^2{\bf r}\! \int_0^1 \!d\alpha \:  
\vert \Psi_{T,L}\,(\alpha,\bfr) \vert ^2 \: \hat\sigma^2\,(x,r^2)\;.
\\
\ee
with the same dipole cross section $\hat\sigma$ as introduced for inclusive
scattering. We account for the $t-$dependence 
by assuming an exponential dependence with the diffractive slope $B_D$.
Thus the $t$-integrated diffractive cross section equals
\be
\label{eq:difft}
\sigma^D(x,Q^2)\;=\;\int_{-\infty}^0 dt\,e^{B_D t}\, 
\left. \frac{d\sigma^D}{d t}\right|_{t=0}\;=\; \frac{1}{B_D}\,
\left.  \frac{d\sigma^D}{d t}\right|_{t=0}\,,
\ee
for both longitudinal and transverse photons.

%\input fig2
%%%%%%%%%%%%%%%%%%%%%%%%%%%%%
\begin{figure}[t] 
  \vspace*{-1cm}
     \centerline{
         \epsfig{figure=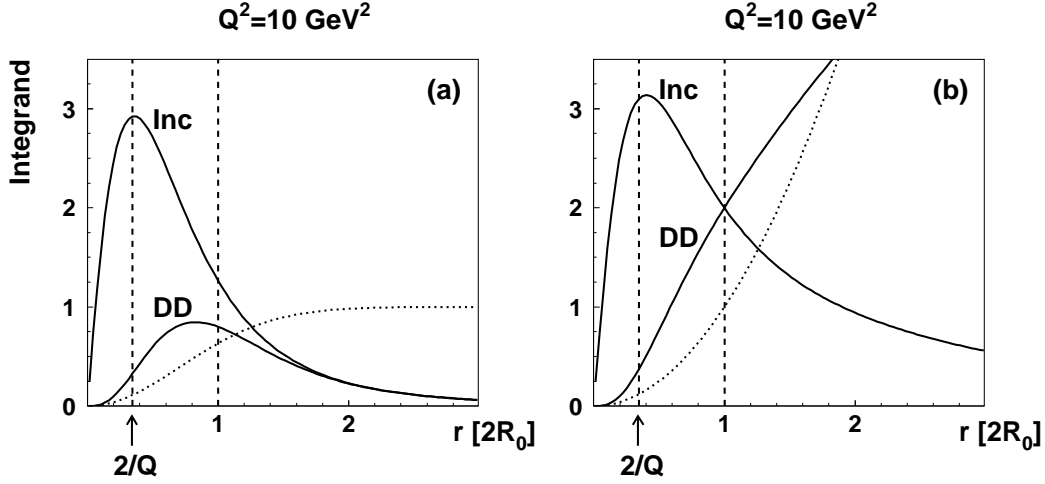,width=15cm}
           }
\vspace*{0.0cm}
\caption{\it The integrands of the inclusive (Inc) and diffractive
(DD) cross sections at $Q^2=10~GeV^2$  
for the following two cases: (a) saturation according to 
Eq.~(\ref{sigmahat}) (dotted line) and (b) no saturation, i.e. 
{\mbox{\rm $\hat\sigma \sim r^2$ }}.\label{rint3}}

\end{figure}
%%%%%%%%%%%%%%%%%%%%%%%%%%%%%

The distributions in $r$ ($q\bar{q}$ dipole size) of the integrand
for inclusive (Eq.~(\ref{eq:1})) and diffractive (Eq.~(\ref{eq:diff1}))
scattering at $Q^2=10~GeV^2$ are shown in Fig.~\ref{rint3}. 
The integrations over $\alpha$ and the azimuthal angle have been performed.
The dotted lines denote the dipole cross section $\hat\sigma$. 
Comparing the two solid lines in Fig.~\ref{rint3}a
we see that for a typical inclusive event 
the main contribution is located around $r \sim 2/Q \ll 2 R_0$.
The diffractive cross section on the other hand 
is dominated by the saturation region $r \sim 2 R_0$.  
The importance of saturation for diffraction is illustrated in Fig.~\ref{rint3}b 
where we let $\hat\sigma$ rise  proportionally to 
$r^2$. While this change has only little effect on the 
inclusive cross section, 
the diffractive cross section becomes strongly divergent 
One, in fact, needs an infrared cutoff - a new, additional scale $R_{cut}$ -
to be introduce by hand. 
As a consequence an additional factor $R_{cut}^2/R_0^2(x)$ 
emerges which leads a result reminiscent of the triple Regge
approach where $\sigma^D \sim x^{-2\lambda}$ instead of 
$\sigma^D \sim x^{-\lambda}$ as we find with saturation.

Fig.~\ref{rint3} also illustrates the idea of Ref.~\cite{Mue}
that diffraction at small $x$ is not a purely soft but semi-hard process.
Let us assume for simplicity that the 'soft regime' begins at  $r=4R_0$.
It becomes quite clear by comparing the two plots in Fig.~\ref{rint3}  
how strongly the soft contribution 
is suppressed due to saturation (blackness of the nucleon). 
The relative fraction of hard contributions ($r<2R_0$) is enhanced
to almost 50\%, making diffractive deep inelastic scattering
a semi-hard process. A related
issue is the smallness of the profile function in central collisions
in $p\bar{p}$ scattering and its consequence for single diffraction.

The following qualitative estimates will help to clarify
the remarks about the importance of saturation for diffraction.
The wave function
in Eq.~(\ref{eq:2}) can be approximated by\footnote{The relation
$K_1(x)\simeq 1/x$ for $x<1$  is used in Eq.~(\ref{eq:2})
in the presented estimation.}
(see also Ref.~\cite{NIK2,JEFF})
\bea\label{align}
\vert \Psi_{T,L}\,(\alpha,\bfr) \vert^2  &\approx&
\frac{6\, \alpha_{em}}{4\,\pi^2 } \, \sum_{f}\,e_f^2 \, \cases  
[\,\alpha^2+(1-\alpha)^2\,]\;\frac{1}{r^2}\;\Theta[\alpha(1-\alpha)Q^2r^2 < 1]
\;\;.
\eea
The leading contribution is associated with the 'aligned jet' configuration.
In the $\gamma^*$-Pomeron CMS the scattering angle $\theta$ is
given by $\cos(\theta)=1-2 \alpha$, i.e. for $\alpha\rightarrow 0\; (1)$
we have $\theta \rightarrow 0 \;(\pi)$. The $\Theta$-function in Eq.~(\ref{align})
enforces the condition that either $\alpha$ or $1-\alpha$ is smaller
than $1/(Q^2 r^2)$. We make use of this condition and the 
$\alpha \leftrightarrow 1-\alpha$ symmetry 
to perform the $\alpha$-integration in Eqs.~(\ref{eq:1}) and (\ref{eq:difft}) and obtain
\bea\label{align2}
\nonumber
\sigma(x,Q^2) &\approx& \frac{6\, \alpha_{em}}{2\,\pi } \, \sum_{f}\,e_f^2
\;\frac{1}{Q^2} 
\;\int_{4/Q^2}^{\infty} \!\frac{d\,r^2}{r^4} 
\;\hat\sigma\,(x,r^2)
\\ 
\\ \nonumber
\sigma^D(x,Q^2) &\approx&
\frac{6\, \alpha_{em}}{32\,\pi^2\,B_D} \, \sum_{f}\,e_f^2\;
\frac{1}{Q^2} 
\int_{4/Q^2}^{\infty} \!\frac{d\,r^2}{r^4}\;
\hat\sigma^2\,(x,r^2) \;\;.
\eea
The lower limit is required, since the factor $1/(Q^2 r^2)$ which results
from the $\alpha$-integration should not exceed 1/4. 
We also approximate the dipole cross section (\ref{sigmahat}) by
\bea\label{align3}
\hat\sigma&\approx& \cases { \sigma_0\;r^2/[4 R_0^2(x)]
&for \hspace*{.5cm}$r^2<4 R_0^2(x)$ 
\cr\cr
\sigma_0 & for \hspace*{.5cm}$r^2>4 R_0^2(x)$}\,
\eea
Inserting (\ref{align3}) into (\ref{align2}) gives after integration
\bea\label{align4}
\nonumber
\sigma(x,Q^2)&\approx&\frac{6\, \alpha_{em}}{2\,\pi } \, \sum_{f}\,e_f^2 
\;\frac{\sigma_0}{4 R_0^2(x)\;Q^2}\;\ln[R_0^2(x)\;Q^2] 
\\
\\ 
\sigma^D(x,Q^2)  &\approx&
\frac{6\, \alpha_{em}}{16\, \pi^2\, B_D} \, \sum_{f}\,e_f^2\;
\frac{\sigma_0^2}{4 R_0^2(x)\;Q^2}
\;\;.\nonumber
\eea
Thus we  obtain an approximate constant ratio of the diffractve
to inclusive cross sections similar to the exact result
in Ref.~\cite{GBW}
\bea\label{ratio}
\frac{\sigma^D}{\sigma}&\approx&\frac{\sigma_0}{8\, \pi\,B_D}
\;\frac{1}{\ln[R_0^2(x)\;Q^2]}\;\;.
\eea
If, on the other hand, we had used 
\bea\label{align5}
\hat\sigma(x,Q^2)&\approx& \sigma_0\;\frac{r^2}{4 R_0^2(x)}
\eea
instead of (\ref{align2}), i.e. no saturation, 
then a cutoff $R_{cut}^2$ would be required leading to
\bea\label{align6}
\nonumber
\sigma(x,Q^2)&\approx&\frac{6\, \alpha_{em}}{2\,\pi } \, \sum_{f}\,e_f^2 
\;\frac{\sigma_0}{4 R_0^2(x)\,Q^2}\;\ln(R_{cut}^2\;Q^2/4) 
\\ 
\\ \nonumber
\sigma^D(x,Q^2) &\approx&
\frac{6\, \alpha_{em}}{32\,\pi^2\,B_D} \, \sum_{f}\,e_f^2\;
\frac{\sigma_0^2\;R_{cut}^2}{[4 R_0^2(x)]^2\,Q^2}\;\;.\nonumber
\eea
The important point is  that the inclusive cross section depends only weakly
on $R_{cut}$ whereas the diffractive cross section
shows a strong dependence. We also realize that under the assumption
(\ref{align5}) the diffractive cross section, being proportional to
$1/R^4_0(x)$, rises at small $x$ twice
as strongly as the inclusive cross section ($\sim x^{-2\lambda}$ as
 mentioned earlier). 
The ratio (\ref{ratio}) would be proportional to
$x^{-\lambda}$ which is clearly not observed at HERA.

To summarize, since the diffractive cross section
is so sensitive to the infrared cutoff which is effectively  given
by $2 R_0(x)$ one can conclude that diffraction directly probes the
transition region. We will now turn to a full description of the diffractive 
deep inelastic scattering data from HERA.

%%%%%%%%%%%%%%%%%%%%%%%%%%%%%%%%%%%%%%%%%%%%%%%%%%%%%%%%%%%%%%%%%%%%%%%%%%%
\section{Diffractive structure function in momentum space representation}

In this section we summarize the relevant contributions to 
the diffractive structure function. 
We use the standard notation for the variables
$\beta=Q^2/(M^2+Q^2)$ and  $x_{\fP}=(M^2+Q^2)/(W^2+Q^2)$ where
$M$ is the diffractive mass and $W$ the total energy of the $\gamma^* p$-process.

Before we start to compute the diffractive structure function it is 
useful to introduce the unintegrated gluon distribution ${\cal F}$
which is related to  the effective dipole
cross section  $(\ref{sigmahat})$ in  the following way
\cite{NIK1,JEFF}:
\bea\label{sig_f}
\hat{\sigma}(x,r)&=&\frac{4 \pi}{3}\;
\int \frac{d^2 \lf_t}{l_t^2} \;
\left[ \;1 - e^{i\;\rf \cdot \lf}\; \right]\; \alpha_s\,{\cal F}(x,l_t^2)\\ 
&=&\frac{4 \pi^2}{3}\;
\int \frac{d l^2_t}{l_t^2} \;
\left[\;1-J_0(l_t r)\;\right]\;\alpha_s\,{\cal F}(x,l_t^2)\;\;.\nonumber
\eea
A short calculation shows that with the following form for $\cal{F}$
\bea \label{f}
\alpha_s\,{\cal F}(x,l_t^2)=\frac{3 \;\sigma_0}{4 \pi^2}
\; R_0^2(x)\;l_t^2\; e^{-R_0^2(x) l_t^2} \;\;,
\eea
one can indeed reproduce Eq.~(\ref{sigmahat}).
At large $Q^2$ the usual gluon distribution $G$ 
can be calculated from ${\cal F}$ by a simple integration:
\bea
\label{gluon1}
x\, G(x,Q^2) &=& 
\int_0^{Q^2}d l_t^2\; {\cal F}(x,l_t^2)\nonumber\\
&=&
\frac{3}{4 \pi^2 \alpha_s}\;\frac{\sigma_0}{R_0^2(x)}\;
\left[1-(1+Q^2R_0^2)\;e^{-Q^2 R_0^2}\right]\\ \nonumber
&\simeq&
\frac{3}{4 \pi^2 \alpha_s}\;\frac{\sigma_0}{R_0^2(x)},
\;\;\;\;\;\;(Q^2 R^2_0(x) \gg 1)
\eea
Important to note is the fact that at large $Q^2$
the gluon distribution exhibits a plain scaling behavior. 
The proper DGLAP evolution in $Q^2$ for the gluon can be added
to our model, e.g. by treating relation (\ref{gluon1}) as the initial distribution
for the linear DGLAP evolution equations. However the results of our
model presented in Fig.~6 suggest that the $Q^2$ dependence of the data at
low $x$ values are properly accounted for in the presented approach and
the additional gluonic evolution will only lead to a moderate improvement.

%The lack of scaling violation is due
%to the absence of the $Q^2$ evolution which should be included at some stage.
%This, however, can only be done at the cost of loosing the simplicity of the
%present model.

\begin{figure}[ht]
\begin{center}
 \epsfig{file=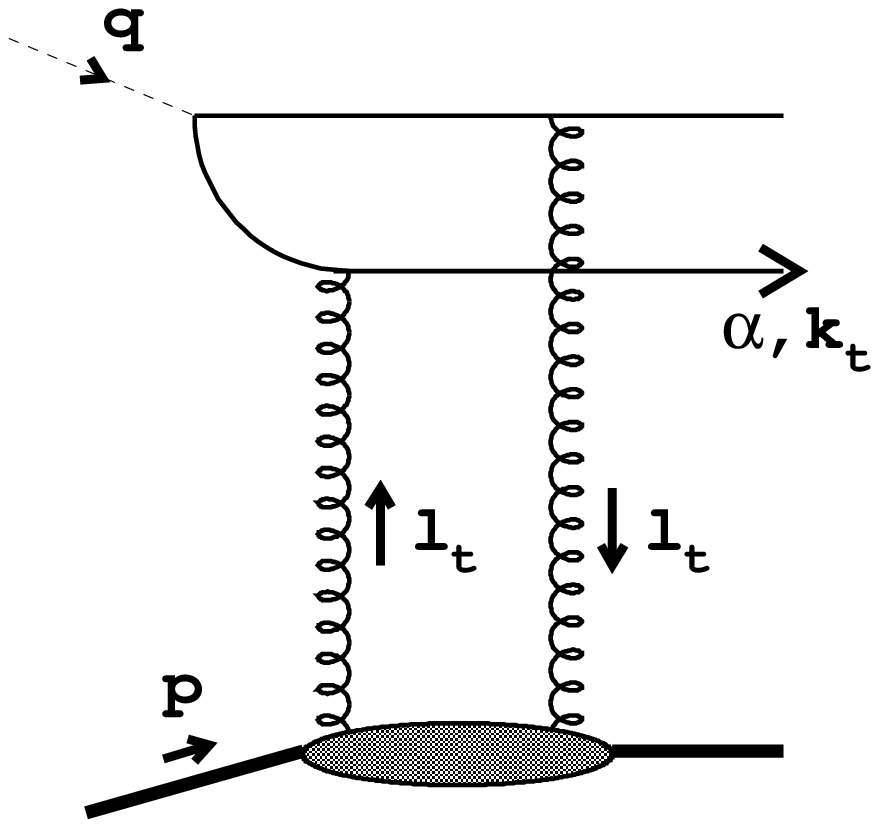,width=8cm}\epsfig{file=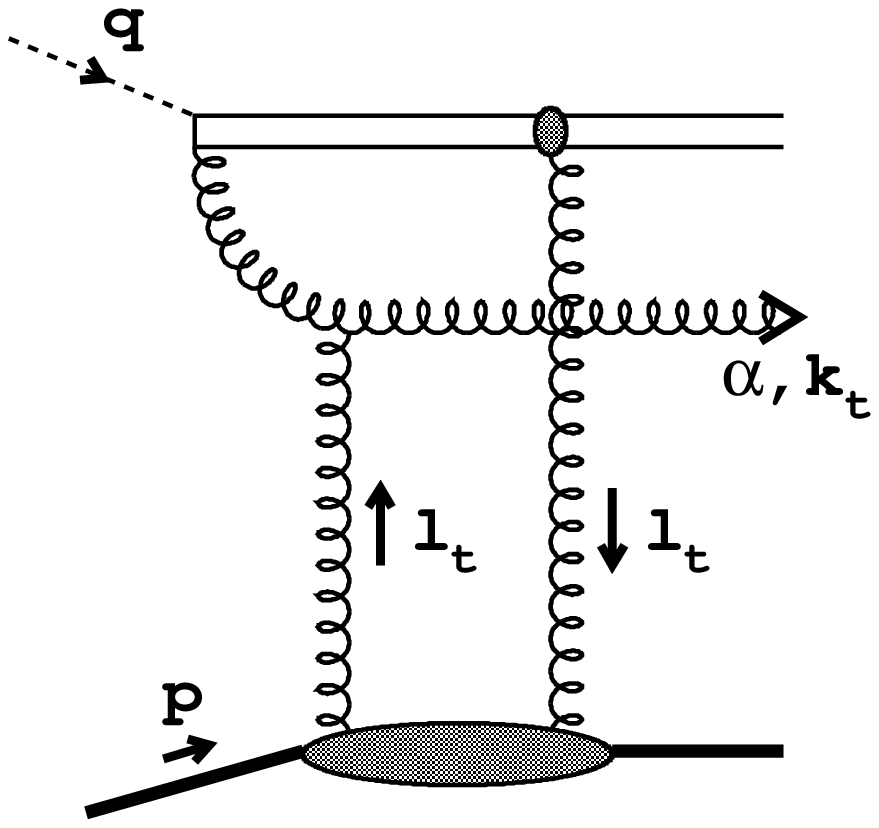,width=8cm}
  \caption{\it Diffractive production of a {\mbox{\rm$ q\bar{q}$}}-pair (left) 
           and the emission of an
           additional gluon (right).}\label{diag}
\end{center}
\end{figure}

We have three terms 
owing to the diffractive production of a quark-antiquark pair
with transverse and longitudinally polarized photons and
the emission of an extra gluon in the final state (Fig.~\ref{diag}). 
The latter contribution 
is only known at present in certain approximations: strong ordering in the
transverse momenta or strong ordering in the longitudinal momentum components.
The first approximation is valid at a very large $Q^2$ and finite diffractive
masses, i.e. finite $\beta$, and picks out the leading logarithm in $Q^2$ 
from the quark box.
The second approximation is valid for large diffractive masses, 
i.e. small $\beta$, and finite $Q^2$ \cite{BJW}. Since the 
diffractive data range around $\beta=0.5$ we will pursue the first
approximation and assume that the transverse momenta of the quarks 
compared to the gluon are much larger.

For a detailed discussion of the derivation of the following formulae
we refer to Ref.~\cite{Wus} and only quote  the result:
\bea\label{diff1}
x_{\fP}F_{\{t,q\bar{q}\}}^D(Q^2,\beta,x_{\fP}) 
&=& \frac{1}{96 B_D}\; \sum_f e^2_f
\;\frac{Q^2}{1-\beta}\;\int_0^1d\alpha\;
[\alpha^2+(1-\alpha)^2] \; \\ &\times&
\left\{\int\frac{dl_t^2}{l_t^2} \alpha_s{\cal F}(x_{\fP},l_t^2)
\left[1-2\beta\;+\;
\frac{l_t^2 - (1-2\beta)\;k^2}{\sqrt{[l_t^2 + k^2]^2
-4 (1-\beta)\;l_t^2 \;k^2}} \right]\right\}^2 \nonumber
\eea
and
\bea\label{diff2}
x_{\fP}F_{\{l,q\bar{q}\}}^D(Q^2,\beta,x_{\fP})
&=& \frac{1}{6 B_D} \; \sum_f e^2_f \;
\int_0^1d\alpha\;k^2\;\beta^2\;\; \\ &\times&
\left\{\int\frac{dl_t^2}{l_t^2} \alpha_s{\cal F}(x_{\fP},l_t^2)
\left[1\;-\;\frac{k^2}
{\sqrt{[l_t^2+k^2]^2 -4 (1-\beta)\;l_t^2\;k^2}} \right]\right\}^2
\nonumber\;\;.
\eea
We have introduced the variable $k^2$ which is defined as
\bea\label{alpha}
k^2&=&\alpha(1-\alpha)\frac{Q^2}{\beta}=\frac{k_t^2}{1-\beta}
\eea
and describes the mean virtuality of the exchange quark in the upper 
part of the diagram. Eq.~(\ref{alpha}) follows from the kinematics 
of the two-body final state. The variable $\alpha$ stems from the Sudakov 
decomposition: $k=\alpha q' + ({|k_t^2|/}{2 \alpha q' \cdot p}) p+k_t$ and
$q'=q+x p$.
The unintegrated structure function ${\cal F}$
is visualized in Fig.~\ref{diag} as the lower blob. It is related to the
inclusive $F_2$ by the optical theorem for zero momentum transfer $t$.
In order to include the $t$-averaged distribution
we have simply divided all expressions by 
the diffractive slope parameter $B_D$ which has to 
be taken from the measurement, see Eq.~(\ref{eq:difft}).

The third contribution takes on the form\footnote{In Ref.~\cite{Wus} 
an overall factor of 2 was miscalculated and needed to be added.}:
\bea\label{diff3}
x_{\fP}F_{\{t,q\bar{q}g\}}^D(Q^2,\beta,x_{\fP})
&=& \frac{9 \;\beta}{64 B_D}\, \sum_f e_f^2 \;\int_0^{Q^2} dk^2\;
\frac{\alpha_s}{2\pi}\;\ln\left(\frac{Q^2}{k^2}\right)\; 
\;\int_\beta^1 \frac{dz}{z^2\,(1-z)^2}\; \nonumber \\
&\times&
\left[\left(1-\frac{\beta}{z}\right)^2
+\left(\frac{\beta}{z}\right)^2\right]
\left\{
\int\frac{dl_t^2}{l_t^2}\;\alpha_s{\cal F}(x_{\fP},l_t^2)
\;\; \right. \\
&\times&\left.\left[z^2+(1-z)^2+\frac{l_t^2}{k^2}-
\frac{[(1-2z)k^2-l_t^2]^2+2z(1-z)k^4}
{k^2\sqrt{(l_t^2+k^2)^2-4(1-z)\;l_t^2\;k^2}}
\right]\right\}^2 \;\;.\nonumber
\eea
In analogy to the previous formulae the variable $k^2$ expresses the mean
virtuality of the exchanged gluon in the upper part of the right diagram
(Fig.~\ref{diag}):
\bea
k^2&=&\frac{k_t^2}{1-z}\;\;.
\eea
The variable $z$ represents the momentum fraction of the upper $t$-channel
gluon with respect to the Pomeron momentum $x_{\fP} p$.
It needs to be stressed that this formula was derived in the spirit
of a leading log($Q^2$) approximation which introduces uncertainties
besides those related to the choice of $\alpha_s$. 
In this approximation the true kinematical
constraints are not exactly fulfilled. The violation of these constraints, 
however, 
gives contributions which are sub-leading in the limit of very large $Q^2$.
An improvement can be achieved by an exact Monte Carlo integration. The
exact treatment of the phase space, however, has to go along with the use
of the exact matrix-element which is not known up to now.
Similar analytic results can be found in Ref.~\cite{HKS}. 
The main difference as compared to our
approach is hidden
in the unintegrated gluon distribution which in Ref.~\cite{HKS} is modeled 
by a heavy quark-antiquark pair. 

There are two limits which are interesting to look at and which
have been discussed in the literature:
the first limit is the triple Regge limit (small $\beta$) in which $z$
can be set to zero in the square bracket of Eq.~(\ref{diff3}).
This leads to
\bea\label{diff4}
x_{\fP}F_{\{t,q\bar{q}g\}}^D(Q^2,\beta,x_{\fP})
&=& \frac{9\; \beta}{64 B_D}\, \sum_f e_f^2 \;\int_0^{Q^2} dk^2\;
\frac{\alpha_s}{2\pi}\;\ln\left(\frac{Q^2}{k^2}\right)\; 
\;\int_\beta^1 \frac{dz}{z^2}\; \nonumber \\
&\times&
\left[\left(1-\frac{\beta}{z}\right)^2
+\left(\frac{\beta}{z}\right)^2\right]
\left\{
\int\frac{dl_t^2}{l_t^2}\;\alpha_s{\cal F}(x_{\fP},l_t^2)
\;\; \right. \\
&\times&\left.2\;\left[\Theta(l_t^2-k^2)+\frac{l_t^2}{k^2}\Theta(k^2-l_t^2)
\right]\right\}^2 \;\;\nonumber
\eea
and agrees with results of Refs.~\cite{BJW,Rys,NIK3}. The other limit is
$l_t^2\ll k^2$ and requires a lower cutoff $k_0^2$ on $k^2$:
\bea\label{diff5}
x_{\fP}F_{\{t,q\bar{q}g\}}^D(Q^2,\beta,x_{\fP})
&=& \frac{9\; \beta}{64 B_D}\, \sum_f e_f^2 \;\int_{k_0^2}^{Q^2} dk^2\;
\frac{\alpha_s}{2\pi}\;\ln\left(\frac{Q^2}{k^2}\right)\; 
\;\int_\beta^1 \frac{dz}{z^2\,(1-z)^2}\; \nonumber \\
&\times&
\left[\left(1-\frac{\beta}{z}\right)^2
+\left(\frac{\beta}{z}\right)^2\right]
\left\{\;\alpha_s x_{\fP} G(x_{\fP},k^2)
\;\; \right. \\
&\times&\left.2\;(1-z)^2\;(1+2z)\;\frac{1}{k^2}\;
\right\}^2 \;\;.\nonumber
\eea 
This result and corresponding approximations for Eqs.~(\ref{diff1}) and 
(\ref{diff2}) have been derived earlier in Ref.~\cite{LeWu}. They have been
utilized in Ref.~\cite{GLM} to perform a similar analysis 
of diffraction as presented in this paper. The approximations used in 
Ref.~\cite{GLM} allow the direct implementation 
of the gluon structure function as given by standard
parameterizations. The result is a too steep increase of the diffractive structure
function with decreasing $x_{\fP}$. The exact formulae in conjunction with
saturation give a much shallower behaviour which is in better agreement with the data
(see below).

%This analysis however concentrates only
%on the region of small diffractive mass  and 
%is based on the expression for the dipol cross section (Eq.~(50) of \cite{GLM})
%which additionally involves integration over the impact parameter. The results
%of the phenomenological analysis presented in \cite{GLM} seems to indicate that
%there is still need for some fine tunning to get better agreement with the
%data.

%%%%%%%%%%%%%%%%%%%%%%%%%%%%%%%%%%%%%%%%%%%%%%%%%%%%%%%%%%%%%%%%%%%%%%%%
\section{Comparison with data}

Before we start our numerical investigation into diffractive scattering
we would like to review the fit to the inclusive data \cite{GBW}. 
The expression for
the structure function $F_2$ we have used in \cite{GBW} 
was derived from Eq.~(\ref{eq:1}) in combination with
the saturation model (\ref{sigmahat})
quoted in Section~2. The parameters were found to be
$\sigma_0 = 23.03~\mbox{\rm mb}$, $\lambda = 0.288$ and $x_0 = 3.04\cdot10^{-4}$.
These parameters enter into the diffractive cross section via the function
${\cal F}$ in Eqs.~(\ref{diff1}-\ref{diff3}). To illustrate
the quality of the fit we plot in Fig.~\ref{fig4} the structure function $F_2(x,Q^2)$ 
in different $Q^2$ bins
in comparison with the data from H1 \cite{H1} and ZEUS \cite{ZEUS} (see also
\cite{GBW} for different comparison). 

The remaining integrations in Eqs.~(\ref{diff1}-\ref{diff3})
have been performed numerically. We consider three light flavors
and assume the diffractive slope parameter $B_D = 6~GeV^{-2}$ which
is somewhat lower than the reported value of $7.1~GeV^{-2}$ \cite{ZEUS-LPS}.
One has, however, to take into account some corrections 
due to double dissociation
(dissociation of the proton) which can be roughly estimated by lowering
the diffractive slope from 7.1 to 6~$GeV^{-2}$.
The coupling constant is kept fixed: $\alpha_s=0.2$.

Fig.~\ref{fig5} shows our result for 
the diffractive structure function $x_{\funp} F^D(x_{\funp},\beta,Q^2)$
at fixed $x_{\funp}=0.0042$ plotted over $\beta$ for various $Q^2$
together with data from ZEUS \cite{ZEUSdif}. Fig.~\ref{fig5.h1} 
contains similar
plots with H1-data for fixed $x_{\funp}=0.003$ \cite{H1dif}. 
The three contributions (\ref{diff1}), (\ref{diff2}) and (\ref{diff3})
have been displayed separately in Fig.~\ref{fig5}. 
The important feature is the 
separation in three distinct regimes of small, medium and high 
$\beta$ where the production of
$q\bar{q}g$, $q\bar{q}$ with transverse and 
$q\bar{q}$ with longitudinally
polarized photons, respectively,  is dominant. It was already
argued in Ref.~\cite{BEKW} that this behavior is mainly due to the nature
of the wave functions rather than the model we use. The relative strength
of the three contributions is fixed by QCD-color factors. The overall
 normalization,
however, directly results from the saturation model without any fits to diffractive data.
This fact is  important to point out,
since in Ref.~\cite{BEKW} the overall and the relative normalization for the mentioned
three contributions was fitted.
One should note that there is no hard gluon component present in our approach
(compare the analyses based on the concept of the `soft' Pomeron structure function
\cite{H1dif,soft}).

The prediction of the $x_{\fP}$-dependence, besides  the overall normalization, 
is an important consequence of the saturation model.
In Fig.~\ref{fig6} and Fig.~\ref{fig7} we compare our predictions with  the
data for $x_{\funp} F^D(x_{\funp},\beta,Q^2)$, now analyzed as a function
of $x_{\funp}$ for different values of $\beta$ and $Q^2$. Notice the good
agreement, especially in the region of moderate and large values of
$\beta$ which corresponds to not too large values of the diffractive mass $M$.
We also reproduce the change of the effective Pomeron
intercept $\bar{\alpha}_{\funp(eff)}$ 
as a function of $Q^2$ for different diffractive masses $M$,
see Fig.~\ref{fig9}. The effective  intercept
is related to the logarithmic $x_{\funp}$-slope $n$ of
$x_{\funp} F^D(x_{\funp},\beta,Q^2)$
through the relation: 
$n=1-2\bar{\alpha}_{\funp(eff)}$.
At low masses $M$ where the longitudinal part dominates the slope in $x_{\fP}$
is slightly steeper due to the enhanced longitudinal
part of the cross section.  Using the effective Pomeron intercept
means having incorporated shrinkage in the context of soft Regge phenomenology.
The rise in $Q^2$ is again mainly caused by the longitudinal part. There is,
however, another effect at work which lowers the intercept at small $\beta$.
The $q\bar{q}g$-contribution has a logarithm $\ln(Q^2/k^2)$ which is approximately
equal to $\ln(Q^2 R_0(x)^2)$. This term effectively lowers the intercept in the regime
where $q\bar{q}g$ dominates, i.e. at small $\beta$.

In  Fig.~\ref{fig8} we show the ratio of the diffractive versus inclusive cross section 
as a function of $W$ for different values of $Q^2$ and the diffractive mass $M$,
in analogy to the analysis in Ref.~\cite{ZEUSdif}. Thus for the presented analysis
we have integrated Eqs.~(\ref{diff1}),~(\ref{diff2}),~(\ref{diff3})  over the 
$\beta-$values
which correspond to the indicated ranges of $M$. The values of the inclusive cross
section were taken from the analysis in Ref.~\cite{GBW}.
The ratio is almost constant
over the entire range of $Q^2$ and $W$ with a slight growth at small $M$
caused by the longitudinal higher twist contribution. One can extract
this behavior directly from the leading twist contributions of Eqs.~(\ref{diff1})
and (\ref{diff3}) by simultaneous rescaling of the $l^2$- and $k^2$-integration 
with respect to $R_0^2$.
We have already discussed that the constant ratio is a particular feature of our
saturation model and certainly deviates from the `conventional' triple Regge approach.

%%%%%%%%%%%%%%%%%%%%%%%%%%%%%%%%%%%%%%%%%%%%%%%%%%%%%%%%%%%%%%%%%%%%%%%%
\section{Diffractive structure function in impact parameter space
representation}

We have started our discussion in  impact parameter space because
it provides a natural way to formulate saturation. 
For this reason we re-derive the formulae of Section 5 in  impact parameter
space. Moreover, the dipole formulation
has its natural foundation in impact parameter space 
\cite{MuePat,BP}. A simple $q\bar{q}$-pair represents an elementary
color dipole which has  an effective scattering 
cross section depending on the separation between the quark
and antiquark.

We will briefly recall the wave function description for a
$q\bar{q}$-pair in impact parameter space using
the conventions of Ref.~\cite{IW} where the subscript $(\pm,\pm)$
denotes the photon- and quark helicity (complex notation):
\bea\label{wavevirtual}
\psi_{(+,+)}(\rf,\alpha)&=& \frac{\sqrt{2}\; i\;e}{2 \pi}\;\alpha \;
\sqrt{\alpha(1-\alpha) Q^2}\;K_1(\sqrt{\alpha(1-\alpha) Q^2 r^2})\;
\frac{\rf}{r}
\nonumber \\
\psi_{(+,-)}(\rf,\alpha)&=& \frac{\sqrt{2}\;i\;e}{2 \pi}\;(1-\alpha)\;
\sqrt{\alpha(1-\alpha) Q^2}\;K_1(\sqrt{\alpha(1-\alpha) Q^2 r^2})\;
\frac{\rf}{r}
\\
\psi_{(-,+)}(\rf,\alpha)&=& \frac{\sqrt{2}\;i\;e}{2 \pi}\;(1-\alpha)\;
\sqrt{\alpha(1-\alpha) Q^2}\;K_1(\sqrt{\alpha(1-\alpha) Q^2 r^2})\;
\frac{\rf^*}{r}
\nonumber \\
\psi_{(-,-)}(\rf,\alpha)&=& \frac{\sqrt{2}\;i\;e}{2 \pi}\;\alpha\;
\sqrt{\alpha(1-\alpha) Q^2}\;K_1(\sqrt{\alpha(1-\alpha) Q^2 r^2})\;
\frac{\rf^*}{r}\nonumber\;\;.
\eea
$K_1$ is the MacDonald-Bessel function, and the variable $\rf$ is  
conjugate to $\kf_t$, i.e. 
\bea
\psi(\rf,\alpha)&=&\int \frac{d^2 \kf_t}{(2\pi)^2} \;e^{i\;\kf_t \cdot \rf}\;
\psi(\kf_t,\alpha)\;\;.
\eea
The longitudinal wave function reads:
\bea
\psi_{(0,\pm)}(\rf,\alpha)&=& \frac{e}{\pi}\;
\alpha(1-\alpha)\; Q\;K_0(\sqrt{\alpha(1-\alpha) Q^2 r^2})\;\;.
\eea

The $\beta$-integrated diffractive structure function can now be
readily expressed in terms of the above wave function and the effective dipole
cross section $\hat{\sigma}$ \cite{JEFF,NIK2}:
\bea\label{impact4}
F_{\{t,q\bar{q}\}}^D(Q^2,x) &=& 
\frac{3 Q^2}{128 \pi^5 B_D}\;\sum_f e_f^2\;
\int_0^1d\alpha\;[\alpha^2+(1-\alpha)^2]\;\; \\
&\times&\alpha(1-\alpha)\; Q^2\;\int d^2\rf\;
K_1^2(\sqrt{\alpha(1-\alpha) Q^2 r^2})\;
\hat{\sigma}^2(r,x)\nonumber
\eea
and
\bea\label{impact5} 
F_{\{l,q\bar{q}\}}^D(Q^2,x) &=& 
\frac{3 Q^2}{32 \pi^5 B_D}\;\sum_f e_f^2\;
\int_0^1d\alpha\;\alpha(1-\alpha)\;\; \\
&\times&\alpha(1-\alpha)\; Q^2\;\int d^2\rf\;
K_0^2(\sqrt{\alpha(1-\alpha) Q^2 r^2})\;
\hat{\sigma}^2(r,x)\;\;.\nonumber
\eea
These two equation demonstrate the simplification one achieves in 
impact parameter space provided the distributions are totally integrated.
They have already been quoted in Eq.~(\ref{eq:diff1}) rewritten 
as diffractive cross section.
The disadvantage, however, is that for differential distributions which depend
on final state energies one has to transform back to momentum space as
in the case of the $\beta$-dependent structure function 
\bea\label{impact6}
x_{\fP}F_{\{t,q\bar{q}\}}^D(Q^2,\beta,x_{\fP})
&=&
\frac{3}{64 \pi^5 B_D}\;\sum_f e^2_f \;\frac{\beta^2}{(1-\beta)^3}\;
\int \frac{d^2\kf_t}{(2\pi)^2}\;\;k_t^4\;\; \nonumber\\
&\times& \frac{1-\frac{2\beta}{1-\beta}\frac{k_t^2}{Q^2}}
{\sqrt{1-\frac{4\beta}{1-\beta}\frac{k_t^2}{Q^2}}}
\;\Theta \left(1-\frac{4\beta}{1-\beta}\frac{k_t^2}{Q^2}\right)\;\; \\
&\times&\int d^2\rf\;\int d^2\rf'\;\;e^{i\;\kf_t \cdot (\rf-\rf')}
\;\;\hat{\sigma}(r,x_{\fP})\;
\hat{\sigma}(r',x_{\fP})\;\; \nonumber\\
&\times&\frac{\rf \cdot \rf'}{r\; r'}
\;K_1\left(\sqrt{\frac{\beta}{1-\beta}k_t^2 r^2}\right)
\;K_1\left(\sqrt{\frac{\beta}{1-\beta}k_t^2 r'^2}\right)\;.\nonumber
\eea
We have made use of Eq.~(\ref{alpha}) to substitute $\alpha$ by $\beta$
keeping $k_t$ fixed. For the longitudinally polarized photons we find
\bea\label{impact8}
x_{\fP}F_{\{l,q\bar{q}\}}^D(Q^2,\beta,x_{\fP})
&=&
\frac{3}{16 \pi^5 B_D}\;\sum_f e^2_f \;\frac{\beta^3}{(1-\beta)^4}\;
\int \frac{d^2\kf_t}{(2\pi)^2}\;\;k_t^4\;\; \nonumber\\
&\times& \frac{{k_t^2}/{Q^2}}
{\sqrt{1-\frac{4\beta}{1-\beta}\frac{k_t^2}{Q^2}}}
\;\Theta \left(1-\frac{4\beta}{1-\beta}\frac{k_t^2}{Q^2}\right)\;\; \\
&\times&
\int d^2\rf\;\int d^2\rf'\;\;e^{i\;\kf_t \cdot (\rf-\rf')}
\;\;\hat{\sigma}(r,x_{\fP})\;
\hat{\sigma}(r',x_{\fP})\;\; \nonumber\\
&\times&
\;K_0\left(\sqrt{\frac{\beta}{1-\beta}k_t^2 r^2}\right)
\;K_0\left(\sqrt{\frac{\beta}{1-\beta}k_t^2 r'^2}\right)\;.\nonumber
\eea
This contribution is suppressed by an extra power in $Q^2$ and therefore is
a higher twist contribution. By using Eq.~(\ref{sig_f}) one can directly
transform Eqs.~(\ref{impact6}) and (\ref{impact8}) into Eqs.~(\ref{diff1})
and (\ref{diff2}).

It should be noted that, when Eqs.~(\ref{impact6}) and (\ref{impact8}) are
integrated over $\beta$, the argument $x_{\fP}$ in $\hat{\sigma}$ is
simply substituted by $x$. This procedure
is valid in the high energy approach as long as the dominant 
contribution is not concentrated at small $\beta$. 
The $\beta$-integration then leads from 
Eqs.~(\ref{impact6}) and (\ref{impact8}) back to 
Eqs.~(\ref{impact4}) and (\ref{impact5}). 
In the case of a gluon in the final 
state one can no longer do a simple substitution but has to integrate
the argument of $\hat{\sigma}$ explicitly.

We will discuss the impact parameterization of the $q\bar{q}g$-final state
in more detail. Our starting point is the wave function for the effective
gluon dipole as described in \cite{Wus}
(we use in this case the vector notation $\mbox{\bf k}_t=(k_t^1,k_t^2)$ 
and $m,n=1,2$
\bea\label{gluondipole}
\psi^{mn}(\alpha,\fk_t)&=&\frac{1}{\sqrt{\alpha(1-\alpha)Q^2}}\;
\frac{k_t^2\;\delta^{mn}\;-\;2\;\fk_t^m\fk_t^n}{k_t^2+\alpha(1-\alpha)Q^2}\\
&=&\frac{1}{\sqrt{\alpha Q^2}}\;
\frac{k_t^2\;\delta^{mn}\;-\;2\;\fk_t^m\fk_t^n}{k_t^2+\alpha Q^2} \nonumber
\eea
The second line of the previous equation is a consequence of the strong 
ordering condition which implies $\alpha\ll 1$. The variable $\alpha$
has been introduced in analogy to Eq.~(\ref{alpha}) and is identical to
${z k_t^2}/{(1-z)Q^2}$,
\bea
\alpha Q^2 &=& \frac{z k_t^2}{1-z}\;=\;z k^2\;\;,
\eea
where $k_t$ is the gluon transverse momentum in this
case and $k^2$ describes the mean virtuality 
of the gluon in the upper $t$-channel.

The following relation illuminates the use of the wave function in momentum
space. After the integration over the azimuth angle of $l_t$ one arrives at
the core expression of Eq.~(\ref{diff3}) 
\bea\label{gluon}
&&\int \frac{d^2\fl_t}{\pi l_t^2}\; 
\alpha_s{\cal F}(x_{\fP},l_t^2)\;\left[
2\;\psi^{mn}(\alpha,\fk_t)\;-\;\psi^{mn}(\alpha,\fk_t+\fl_t)\;-\;
\psi^{mn}(\alpha,\fk_t-\fl_t)\right] 
\\ \nonumber
\\ \nonumber
&=&\int \frac{dl_t^2}{l_t^2}\;
\frac{ \alpha_s{\cal F}(x_{\fP},l_t^2)}{\sqrt{\alpha Q^2}}
\left[1 - \frac{2k_t^2}{k_t^2+\alpha Q^2} -
\frac{l_t^2}{k_t^2} - \frac{\alpha Q^2}{k_t^2}\right.\\
&& \left.\hspace{4cm}+\; 
\frac{[l_t^2-k_t^2+\alpha Q^2]^2+2k_t^2\alpha Q^2}{k_t^2
\sqrt{[l_t^2+k_t^2+\alpha Q^2]^2 -4l_t^2k_t^2}}\right]
\left\{2\frac{\fk_t^m \fk_t^n}{k_t^2} - \delta^{mn}\right\}\nonumber\\
&=&\int\frac{dl_t^2}{l_t^2}\;\nonumber
\frac{\alpha_s{\cal F}(x_{\fP},l_t^2)}{(1-z)\sqrt{z k^2}}
\left[z^2+(1-z)^2+\frac{l_t^2}{k^2} \right. \nonumber\\
&& \left.\hspace{4cm}-\;\frac{[(1-2z)k^2-l_t^2]^2+2z(1-z)k^4}
{k^2\sqrt{(l_t^2+k^2)^2-4(1-z)\;l_t^2\;k^2}}\right]
\left\{\delta^{mn}-2\frac{\fk_t^m \fk_t^n}{k_t^2}\right\}
\;\;.\nonumber
\eea
The four terms $\psi^{mn}(\alpha,\fk_t)+\psi^{mn}(\alpha,\fk_t)-
\psi^{mn}(\alpha,\fk_t+\fl_t)-\psi^{mn}(\alpha,\fk_t-\fl_t)$ represent the four
possible ways of coupling the two $t$-channel gluons to the gluon dipole
(without crossing in the $t$-channel).
The Fourier transformation of the wave function leads to
\bea\label{psi_r}
\psi^{mn}(\alpha,\fr)
&=&-\frac{1}{2\pi}\left(\delta^{mn}-2\frac{\rf^m \rf^n}{r^2}\right)\;
\sqrt{\alpha Q^2}\;K_2(\sqrt{\alpha Q^2 r^2})\;\;.
\eea
Inserting the Fourier transform into the first line of Eq.~(\ref{gluon})
and using Eq.~(\ref{sig_f})  we find
\bea
&&\int \frac{d^2\fl_t}{\pi l_t^2}\; 
\alpha_s{\cal F}(x_{\fP},l_t^2)\;\left[
2\;\psi^{mn}(\alpha,\fk_t)\;-\;\psi^{mn}(\alpha,\fk_t+\fl_t)\;-\;
\psi^{mn}(\alpha,\fk_t-\fl_t)\right]\nonumber\\
&=&\int d^2\fr \;\psi^{mn}(\alpha,\fr)\; \;e^{i\;\kf_t \cdot \rf}\;
\int \frac{d^2\fl_t}{\pi l_t^2}\; \alpha_s{\cal F}(x_{\fP},l_t^2)\;
\left(2-e^{i\;\lf_t \cdot \rf}-e^{-i\;\lf_t \cdot \rf}\right)\\
&=&2\;\int d^2\fr \;\psi^{mn}(\alpha,\fr)\; \;e^{i\;\kf_t \cdot \rf}\;
\frac{3}{4 \pi^2}\;\hat{\sigma}(x_{\fP},r)\nonumber\;\;.
\eea
We can now rewrite Eq.~(\ref{diff3}) in impact parameter space as\footnote{
A missing factor $1/2$ in the journal version was inserted in this update.}
\bea\label{impact_gluon}
x_{\fP}F_{\{t,q\bar{q}g\}}^D(Q^2,\beta,x_{\fP})
&=&
\frac{81\; \beta}{512 \pi^5 B_D}\;\sum_f e_f^2 \;\frac{\alpha_s}{2\pi}
\;\int_\beta^1 \frac{dz}{z}\;\left[\left(1-\frac{\beta}{z}\right)^2
\;+\;\left(\frac{\beta}{z}\right)^2\right]\;
\frac{z}{(1-z)^3}\; \nonumber\\
&\times&
\int\frac{d^2\fk_t}{(2\pi)^2}
\;k_t^4\;\ln\left(\frac{(1-z)Q^2}{k_t^2}\right)\;
\Theta\left[(1-z)Q^2-k_t^2\right]
\;\;\\
&\times&
\int d^2\rf\;\int d^2\rf'\;\;e^{i\;\kf_t \cdot (\rf-\rf')}
\;\;\hat{\sigma}(r,x_{\fP})\;\hat{\sigma}(r',x_{\fP})\;
\left(\delta^{mn}-2\frac{\rf^m \rf^n}{r^2}\right)\;\;\nonumber\\
&\times&
\left(\delta^{mn}-2\frac{\rf'^m \rf'^n}{r'^2}\right)
\;K_2\left(\sqrt{\frac{z}{1-z}k_t^2 r^2}\right)
\;K_2\left(\sqrt{\frac{z}{1-z}k_t^2 r'^2}\right)\nonumber
\eea
Again, a direct computation of Eq.~(\ref{impact_gluon}) after substituting
$\hat{\sigma}$ according to Eq.~(\ref{sig_f}) reproduces the result
of Eq.~(\ref{diff3}).

The impact parameter representation in Eqs.~(\ref{impact6}) and (\ref{impact_gluon})
demonstrate the similarity of our approach and the semiclassical approach of
Ref.~\cite{BGH}. It suggests that the two-gluon exchange model
can be extended to multi-gluon exchange without changing the basic analytic structure.
The leading color tensors in the limit of large N$_c$ (number of colors)
for a quark- and a gluon-loop
with an arbitrary number of $t$-channel gluons attached to them are
found to be identical up to an overall constant factor \cite{Carlo}. The large
N$_c$ result differs only slightly from N$_c=3$ in the two-gluon
exchange model and, hence, multi-gluon exchange is expected to give very similar
results as the two-gluon exchange.
%%%%%%%%%%%%%%%%%%%%%%%%%%%%%%%%%%%%%%%%%%%%%%%%%%%%%%%%%%%%%%%%%

\section{Conclusions}

In our analysis we successfully describe diffractive 
deep inelastic scattering using 
the saturation model proposed in Ref.~\cite{GBW}. This model reproduces
quite accurately the $\beta$- and $x_{\fP}$- distributions as measured by
H1 and ZEUS \cite{H1dif,ZEUSdif} without tuning or fitting
any additional parameters. 

As demonstrated in Ref.~\cite{GBW} 
saturation naturally explains the transition  of the inclusive structure
function $F_2$ from high to low values of $Q^2$.
Diffractive scattering is even more effected by saturation 
(see Section 3). The constant ratio of the
diffractive versus inclusive cross sections as observed at HERA
is a direct consequence of saturation.
It was also pointed out that soft contributions are significantly suppressed 
leading to a relative enhancement of semi-hard contributions. This fact allows
the conclusion that diffraction in deep inelastic scattering
is a semi-hard process \cite{Mue}. 
The effective Pomeron intercept is higher than
expected from a `soft' Pomeron approach \cite{H1dif,soft}.
The $\beta$-spectrum depends only weakly 
on the model and is therefore more universal. 

The model we choose for saturation is purely phenomenological.
An alternative model without low-$x$ saturation can be found in Ref.~\cite{FKS}.
A completely theoretical framework involves  non-linear QCD evolution equations
as proposed in Refs.~\cite{GLR,Kovner,Kovchegov}.
We believe, however,  that our model represents the basic 
dynamics at very low $x$, since it allows us to describe
a wide range of data in a satisfactory way. 

We can use our analysis to predict diffractive charm production. This requires the  
discussion of factorization, the introduction of diffractive parton distributions
and the evolution of the diffractive final state.
The detailed discussion of these topics will be presented elsewhere \cite{GBW2}.

%%%%%%%%%%%%%%%%%%%%%%%%%%%%%%%%%%%%%%%%%%%%%%%%%%%%%%%%%%%%%%
\vskip 1cm
\noindent {\large \bf Acknowledgements}   
   
We thank H. Abramowicz,  C. Ewerz, J. Forshaw, G. Kerley, J. Kwiecinski, E. Levin,
M. McDermott, G. Shaw and A. Stasto
for useful discussions. H. Kowalski, P. Newman  
kindly provided us with the data from ZEUS and H1. We are particular grateful
to H. Kowalski for his help in preparing Fig.~\ref{fig9} and Fig.~\ref{fig8}.
K.G-B. thanks the Department of Physics
of the University of Durham for hospitality.
This research has also been supported in part by the Polish State    
Committee for Scientific Research grant No.~2~P03B~089~13 and by the EU Fourth    
Framework Programme \lq Training and Mobility of Researchers' Network, \lq    
Quantum Chromodynamics and the Deep Structure of Elementary Particles', contract    
FMRX-CT98-0194 (DG~12-MIHT).   
%%%%%%%%%%%%%%%%%%%%%%%%%%%%%%%%%%%%%%%%%%%%%%%%%%%%%%%%%%%%%
\newpage\newpage
\section*{Appendix}
In this appendix we would like to recall the derivation of Eq.~(\ref{diff3}) which
represents the contribution due to the emission of an additional gluon \cite{Wu_diss}.
We choose light-cone gauge with the gauge fixing condition
$q'\cdot A=0$ ($A$ is the gluon potential, $q'=q+x p$). The frame which 
naturally corresponds
to this choice of gauge is the Breit frame, i.e. the frame
in which the proton is fast moving. All   
quasi-Bremsstrahlungs gluons emitted from the $q\bar{q}$-pair can be neglected.
Those from the incoming partons on the other hand have 
to be taken into account.
 
The polarization vector $\epsilon$ 
for real gluons and the polarization tensor for the gluon 
propagator $d^{\nu\mu}$ read:
\beqn\label{app1}
\epsilon^\mu(k)&=&\epsilon_t^\nu (k)
               -q'^\mu\;\frac{k_t\cdot \epsilon_t(k)}{k\cdot q'}  \\
d^{\nu\mu}(k)&=&g^{\nu\mu}-\frac{k^\nu q'^\mu+q'^\nu k^\mu}{k\cdot q'}
\;\;.\nonumber\
\eeqn
\begin{figure}[ht]
\begin{center}
\epsfig{file=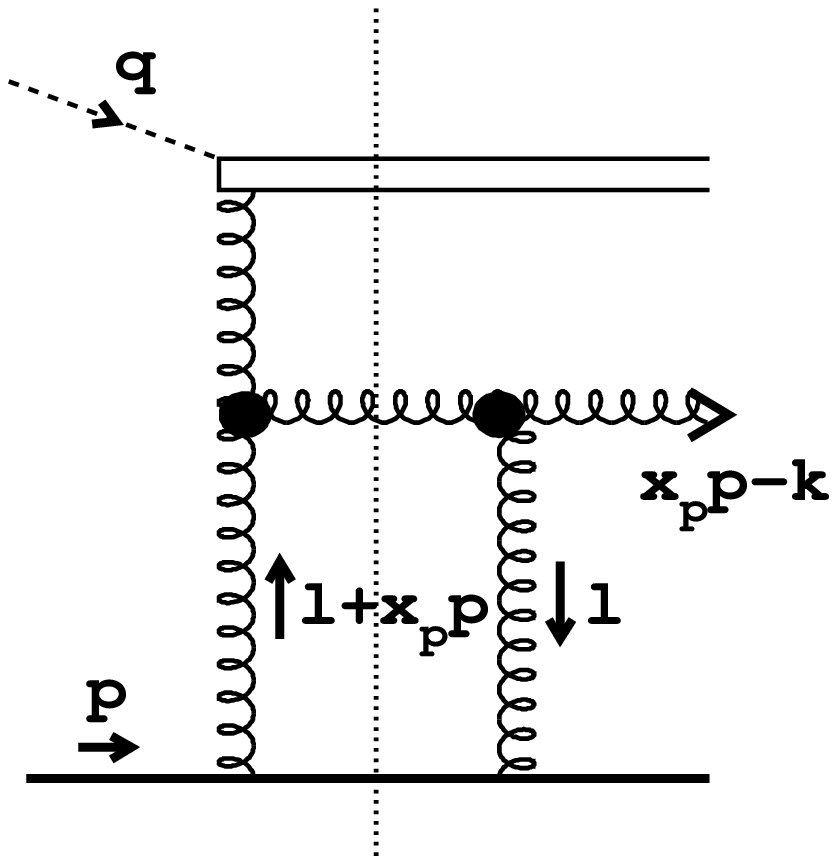,width=5.2cm}
\epsfig{file=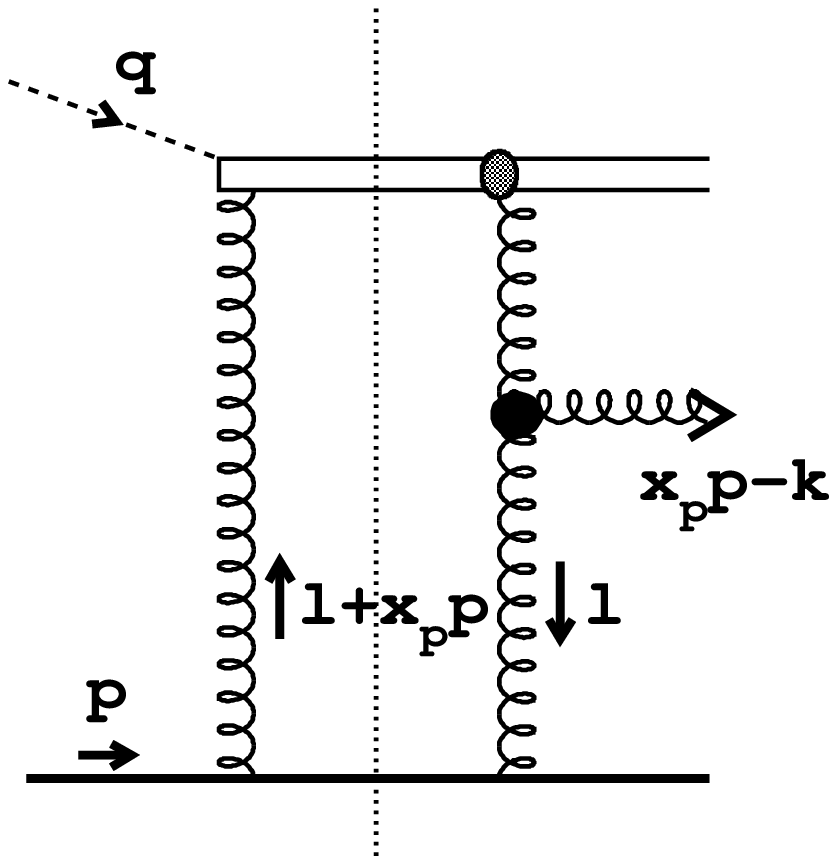,width=5.2cm}
\epsfig{file=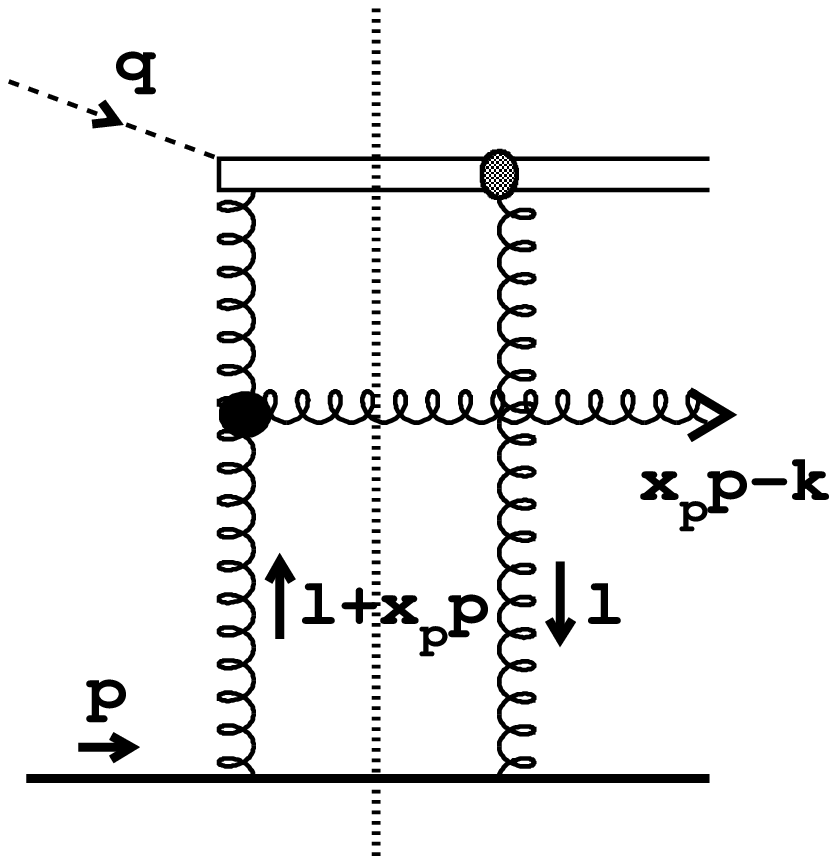,width=5.2cm}
\caption{Gluon radiation.}
  \label{fig_app1}
\end{center}
\end{figure}

\begin{figure}[ht]
\begin{center}
\epsfig{file=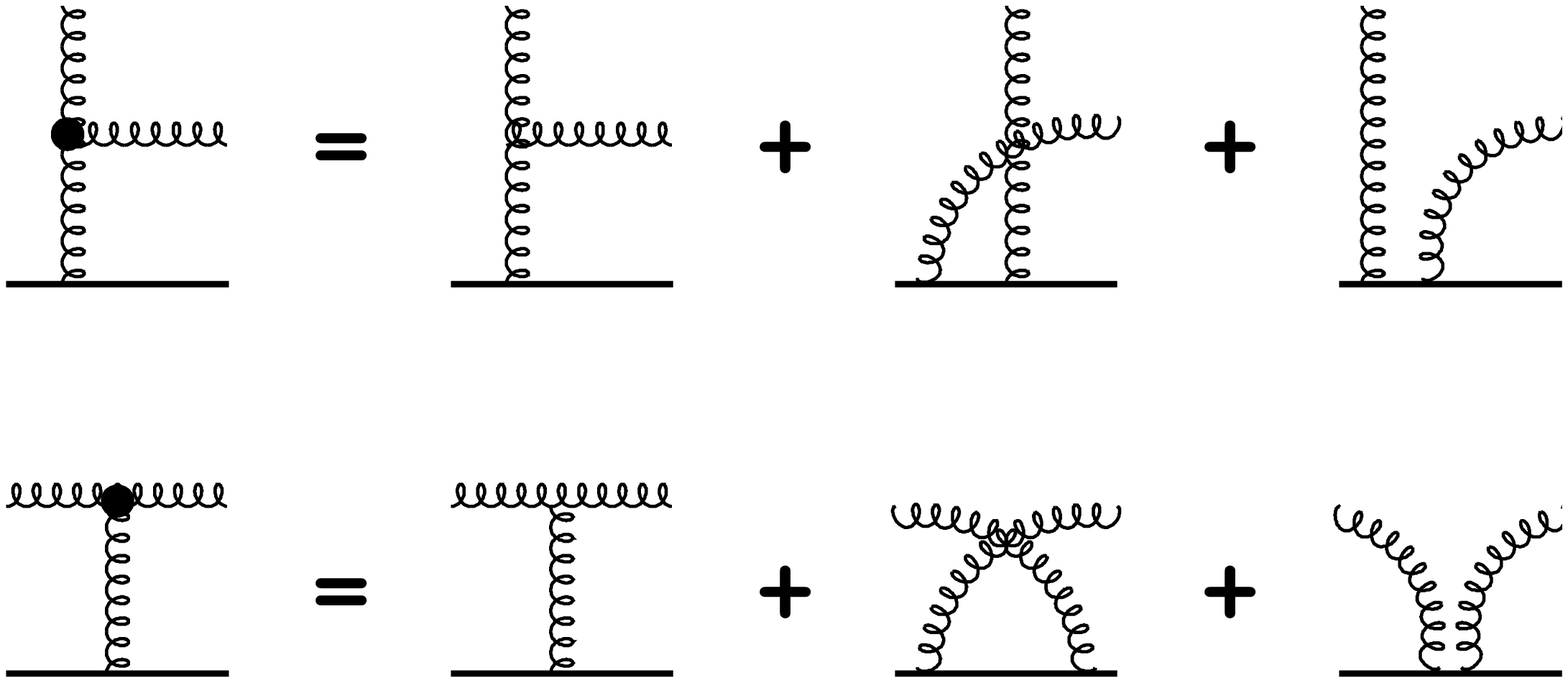,width=13cm}
  \caption{Effective triple gluon couplings.}
  \label{fig_app4}
\end{center}
\end{figure}

\begin{figure}  
   \vspace*{-1cm}  
    \centerline{  
     \epsfig{figure=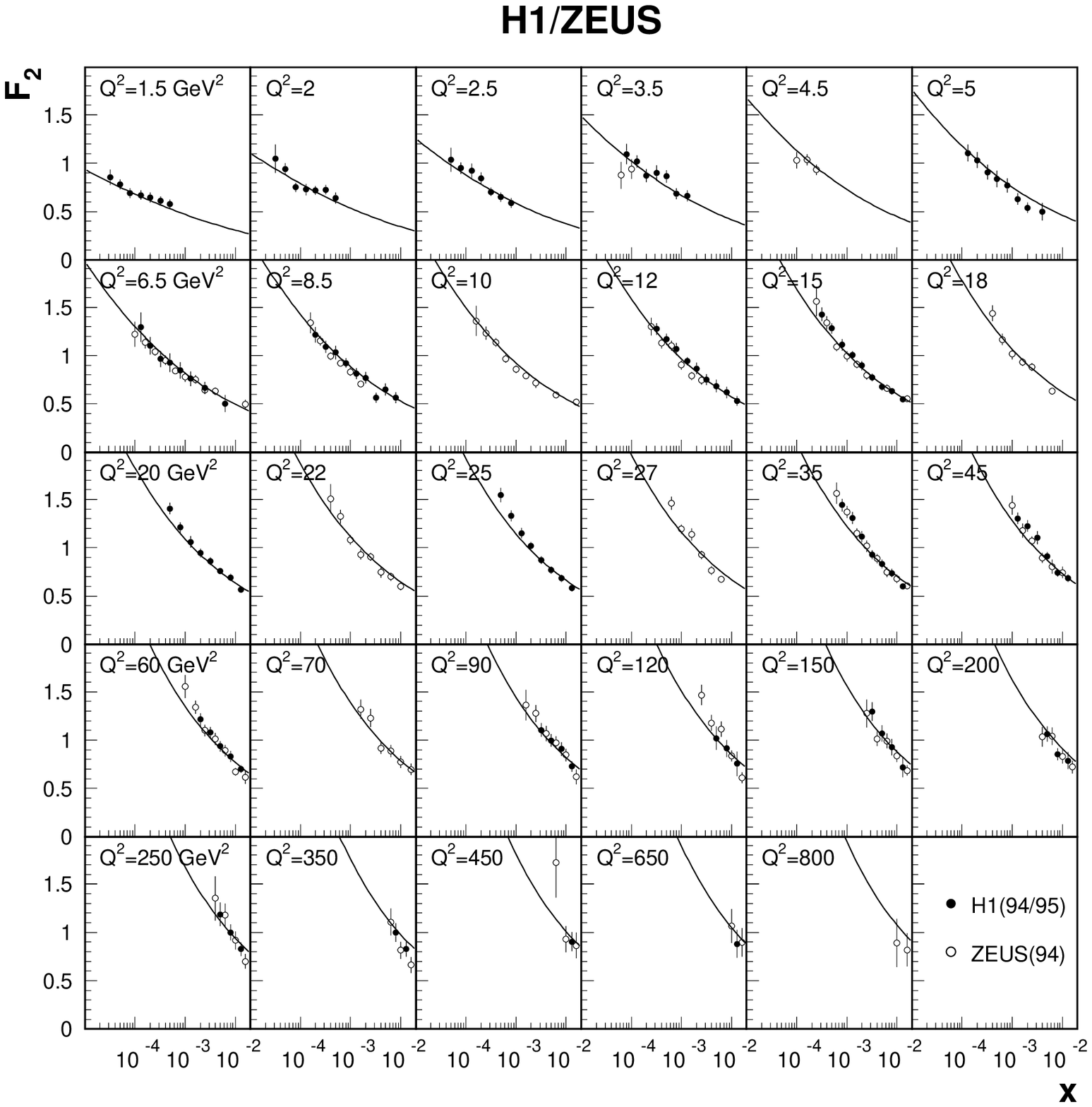,width=18cm}  
               }  
    \vspace*{-0.5cm}  
\caption{The results (solid lines) of the fit to
the inclusive HERA data on $F_2$ for different values of $Q^2$, using the model
of $[1]$ with saturation.  
}  
\label{fig4}  
\end{figure}

\begin{figure} 
   \vspace*{-1cm} 
    \centerline{ 
     \epsfig{figure=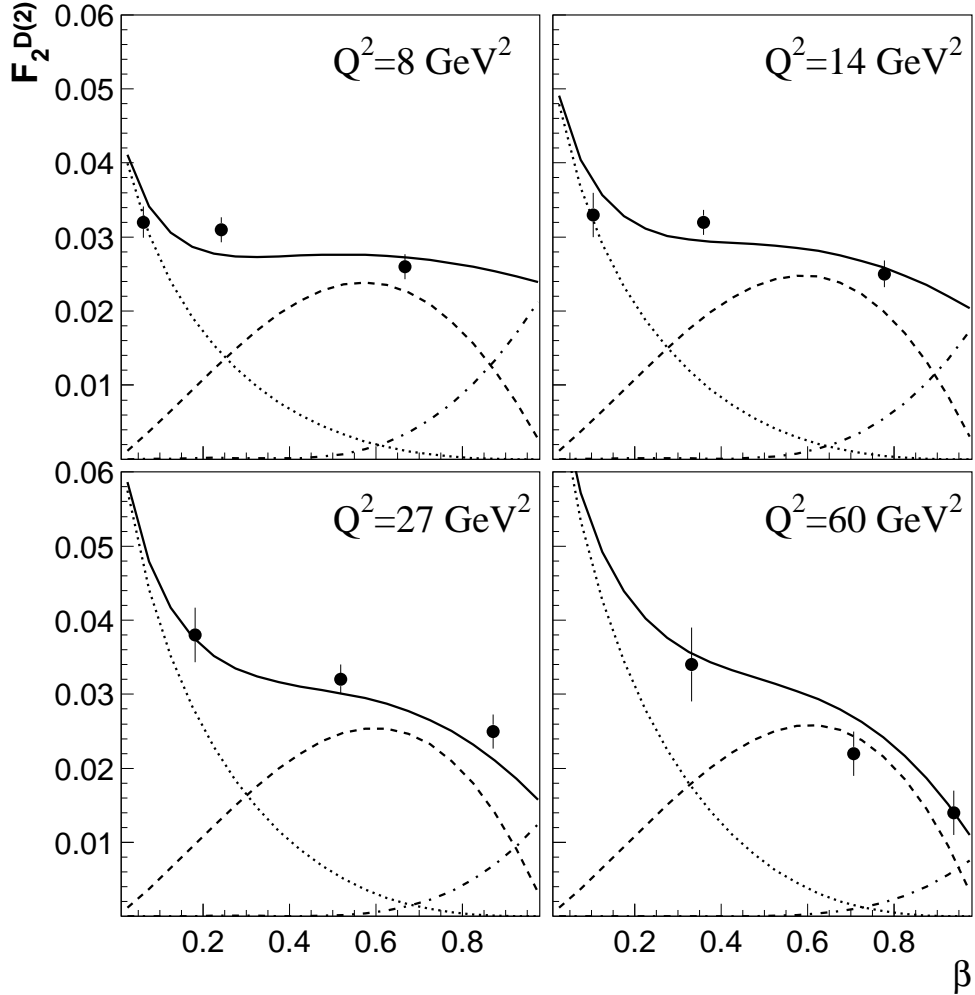,width=15cm} 
               } 
    \vspace*{-0.5cm} 
\caption{The diffractive structure function $x_{\funp} F^D(x_{\funp},\beta,Q^2)$
for $x_{\funp}=0.0042$ as a function of $\beta$. The dashed
lines show the $q\bar{q}$ contribution for transverse photons 
{\mbox{\rm (\ref{diff1})}}, the dot-dashed
lines correspond to the contribution from longitudinal
photons {\mbox{\rm (\ref{diff2})}} and
the  dotted lines illustrate
the $q\bar{q}g$ component {\mbox{\rm (\ref{diff3})}}. 
The solid line is the total contribution and the data are from ZEUS. 
} 
\label{fig5} 
\end{figure}

\begin{figure} 
   \vspace*{-1cm} 
    \centerline{ 
     \epsfig{figure=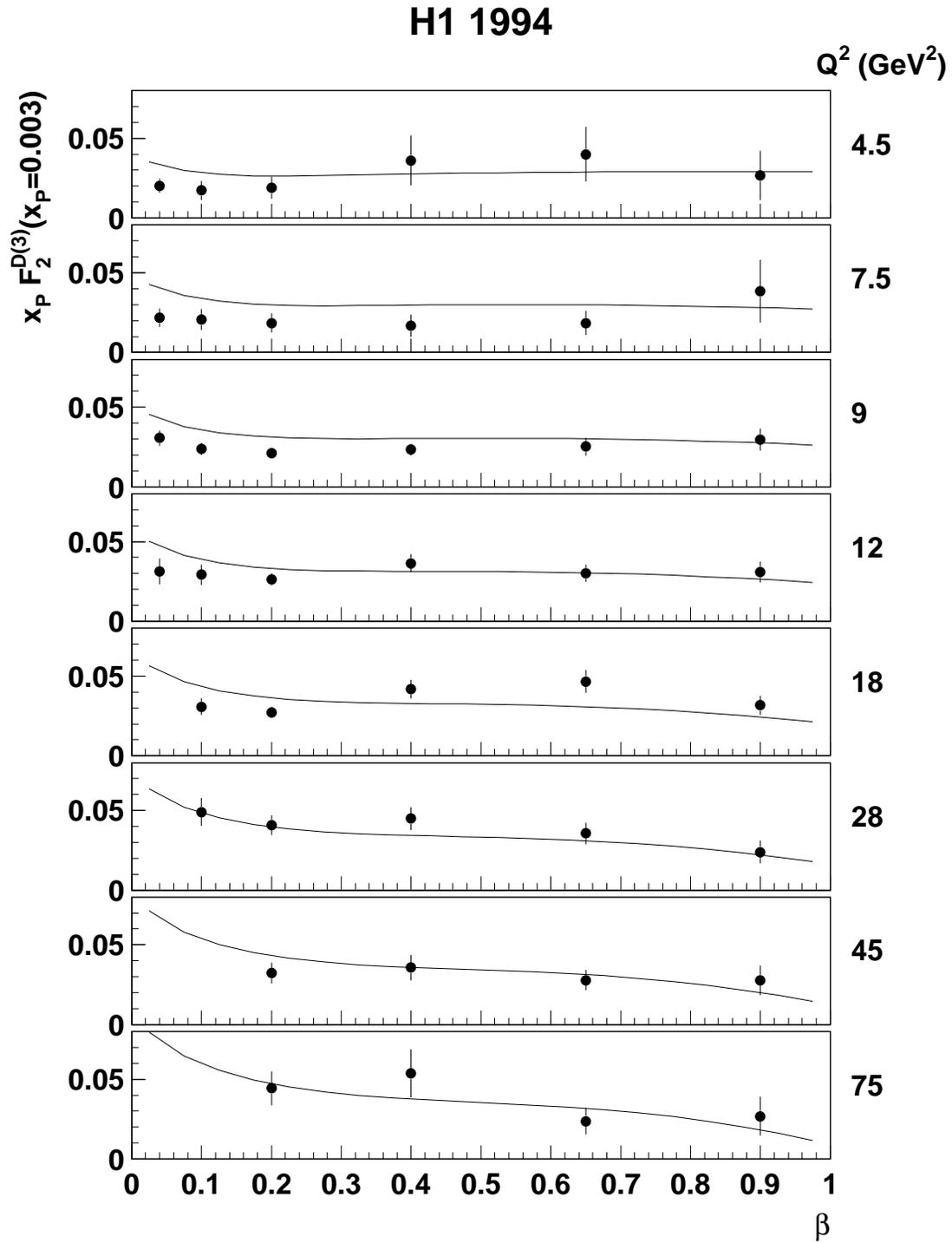,width=15cm} 
               } 
    \vspace*{-0.5cm} 
\caption{The same comparison as in  Fig.~\ref{fig5} 
but with H1 data. Only the total contribution
is shown (solid lines).
} 
\label{fig5.h1} 
\end{figure}

\begin{figure}  
   \vspace*{-1cm}  
    \centerline{  
     \epsfig{figure=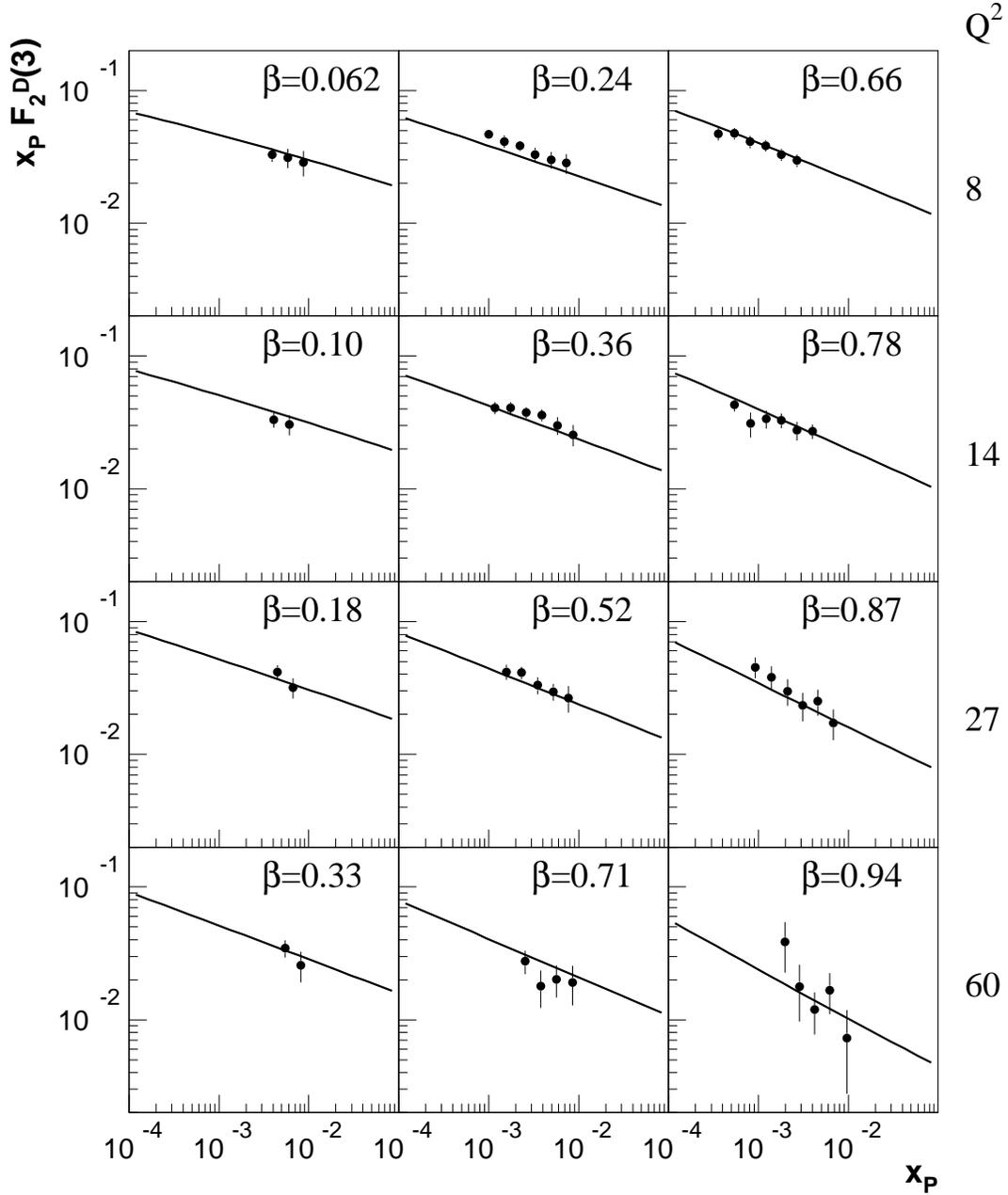,width=15cm}  
               }  
    \vspace*{-0.5cm}  
\caption{The diffractive structure functions $x_{\funp} F^D(x_{\funp},\beta,Q^2)$ 
as measured by ZEUS plotted
as a function of $x_{\funp}$ for different values 
of $\beta$ and $Q^2$ (in units of $GeV^2$).
}  
\label{fig6}  
\end{figure}

\begin{figure}  
   \vspace*{-1cm}  
    \centerline{  
     \epsfig{figure=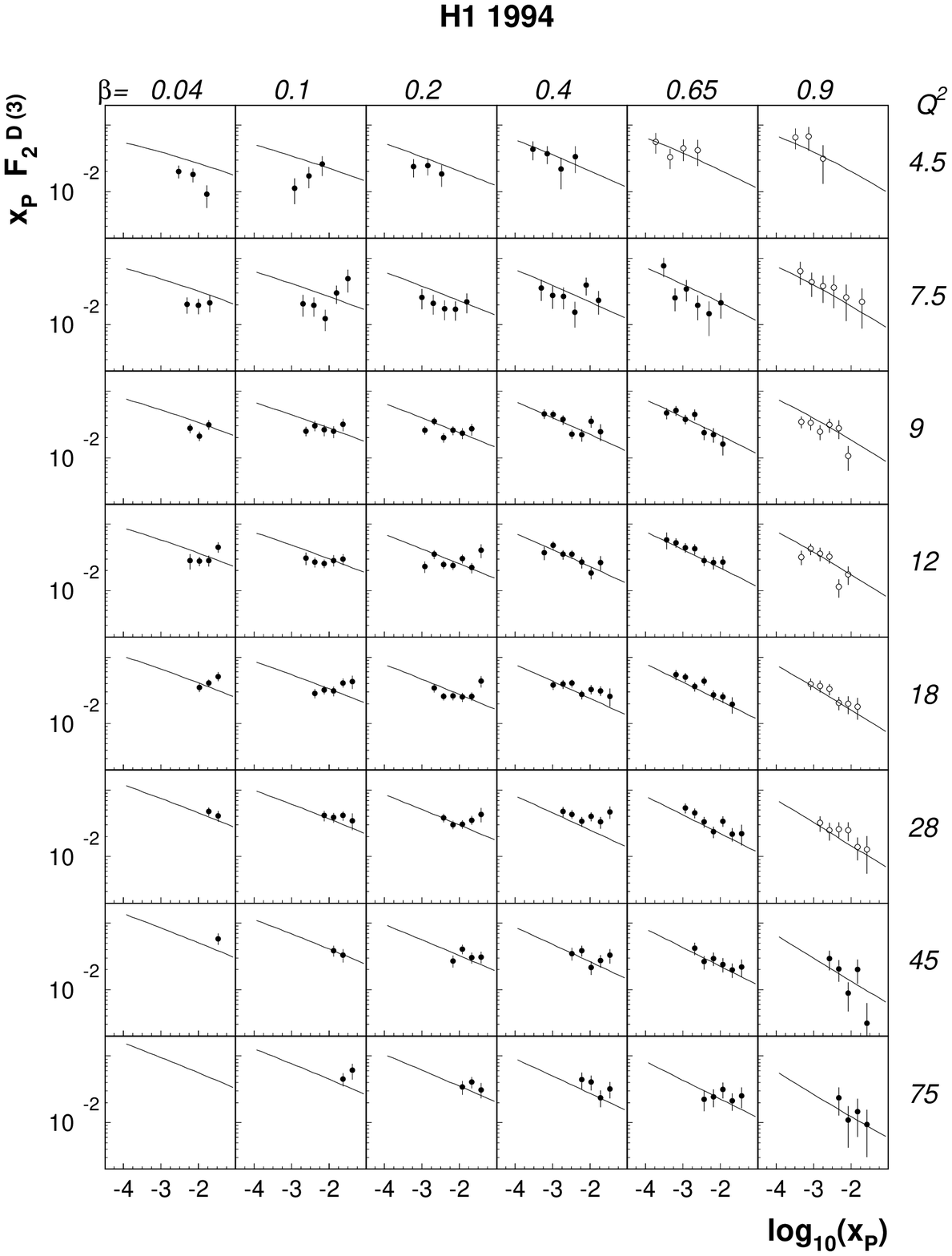,width=15cm}  
               }  
    \vspace*{-0.5cm}  
\caption{The same as in Fig.~\ref{fig6} but for H1
data. $Q^2$ values are in units of $GeV^2$.
}  
\label{fig7}  
\end{figure}

\begin{figure}  
   \vspace*{-1cm}  
    \centerline{  
     \epsfig{figure=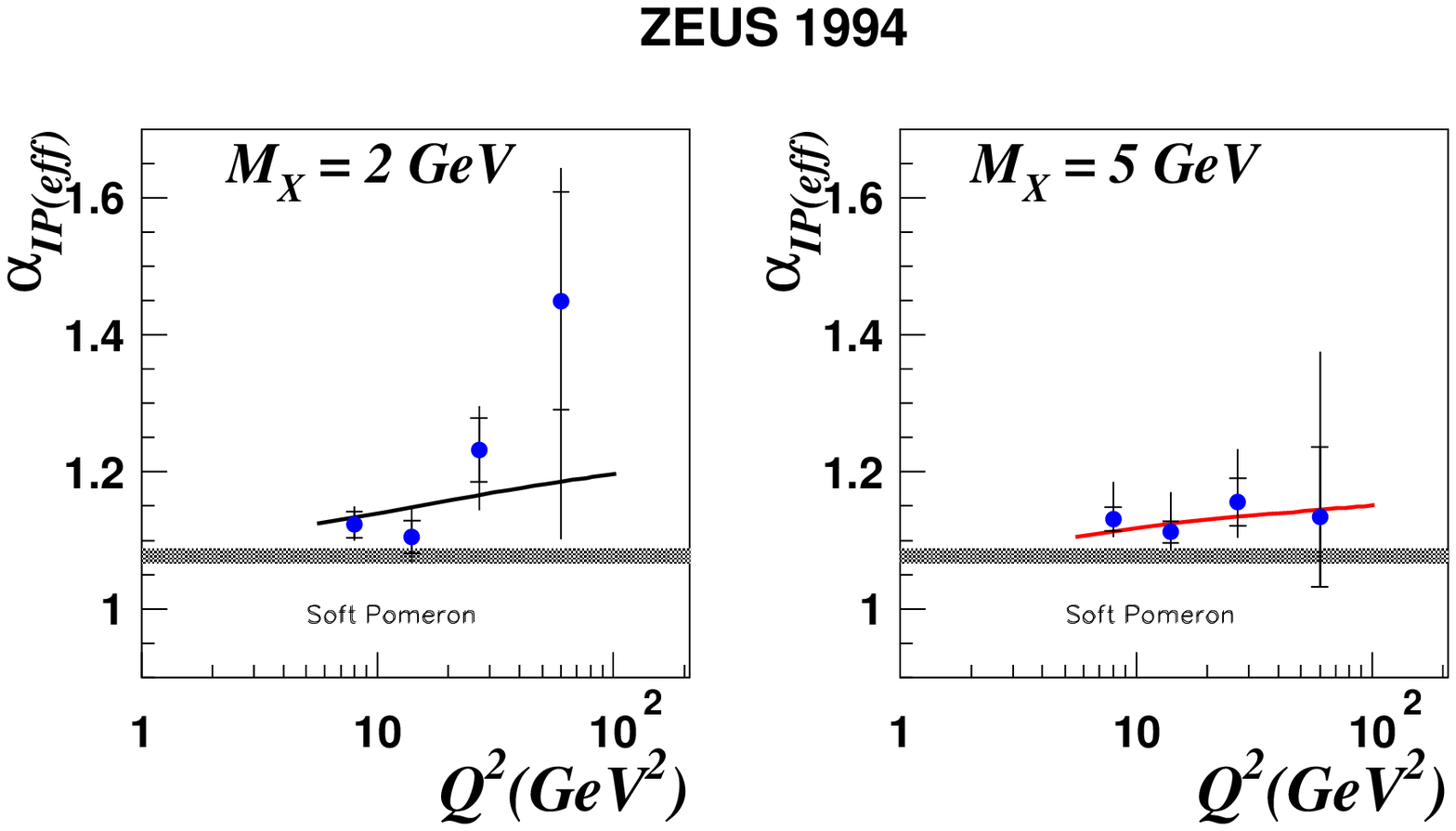,width=15cm}  
               }  
    \vspace*{-0.5cm}  
\caption{The effective Pomeron slope as defined in the text
as a function of $Q^2$ for two values of the diffractive mass $M$.
}  
\label{fig9}  
\end{figure}

\begin{figure}  
   \vspace*{-1cm}  
    \centerline{  
     \epsfig{figure=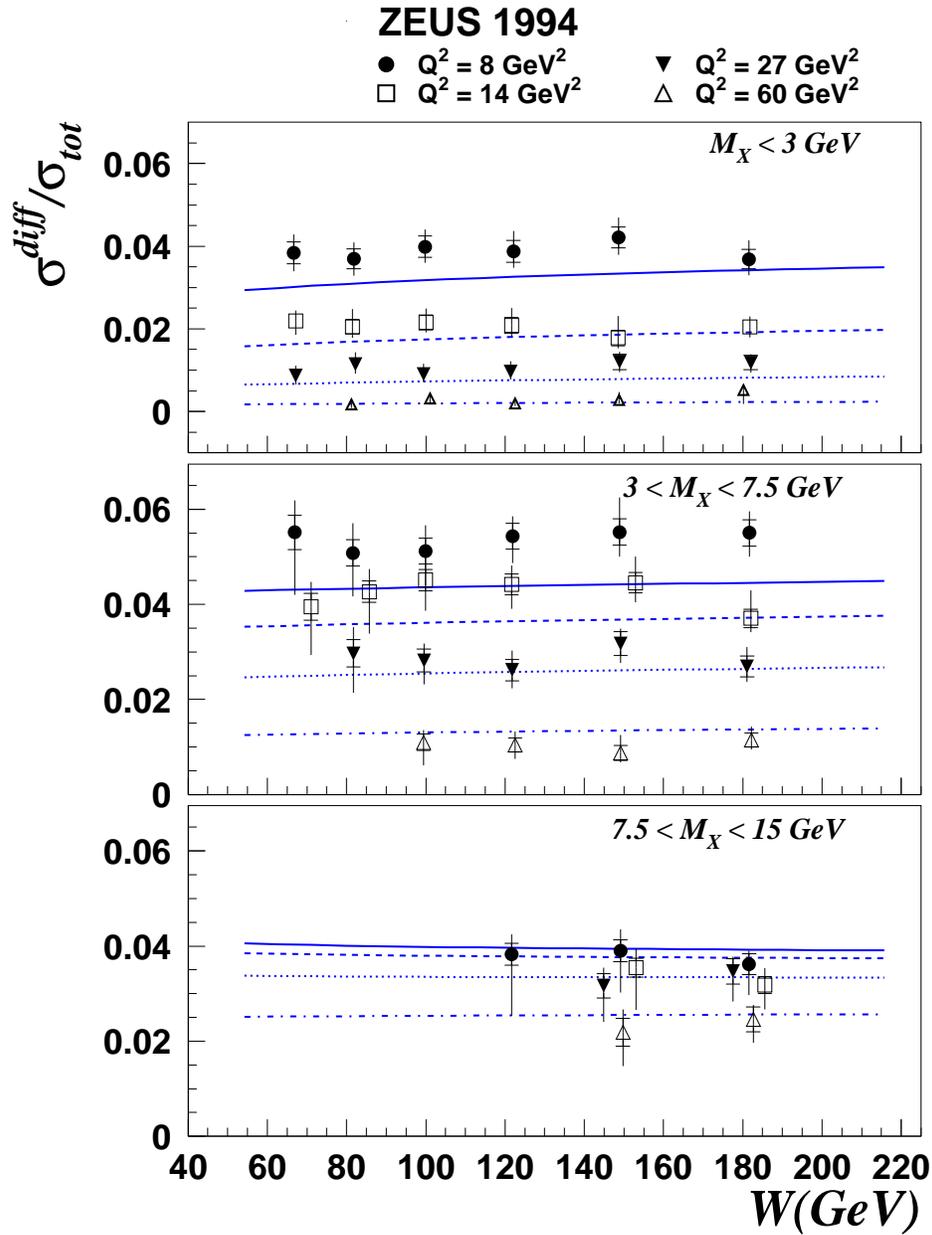,width=15cm}  
               }  
    \vspace*{-0.5cm}  
\caption{The ratio of the diffractive versus the inclusive cross sections
as a function of $W$ for different values of  $Q^2$ and the diffractive mass $M_X$.
}  
\label{fig8}  
\end{figure}

Figure \ref{fig_app1} shows all the essential diagrams. 
The two diagrams to the left have 
a similar momentum structure and will be summed up right from the beginning 
whereas the diagram on the right will be calculated separately.
The bottom line in all the diagrams represents a quark. It is 
accompanied by other 'spectator-quarks' which are not shown explicitly.
The cut through the diagrams effectively
subdivides the whole amplitude into two subprocesses. 
We will introduce effective three gluon couplings
which are the sum of the original three gluon coupling and extra
Bremsstrahlungs contributions (see Fig.~\ref{fig_app4}). These couplings and
their analytic formulae represent the core of the whole calculation. 
The blob at the top of the right $t$-channel gluon in
Fig.~\ref{fig_app1} indicates the simultaneous coupling of the $t$-channel gluon
to the $q\bar{q}$-pair which in color space combines into a gluon.

Before starting the calculation one has to recall and make use of the kinematic
assumptions made in this approach. Firstly, there is the Regge limit with respect
to the lower part of the diagram, i.e. the emitted gluon and the 
quark at the bottom have an invariant subenergy much larger than the diffractive
mass $M$. The high energy assumption
allows one to simplify the $t$-channel propagator as to
\beqn\label{app2}
d^{\rho\sigma}(l+x_{\fP}p)&=&g^{\rho\sigma}
-\frac{(l+x_{\fP}p)^\rho q'^\sigma + q'^\rho (l+x_{\fP}p)^\sigma}
{(\beta_l+x_{\fP})p\cdot q'} \\
&\simeq&-\frac{l_t^\rho q'^\sigma}{(\beta_l+x_{\fP}) p\cdot q'} \nonumber 
\eeqn
where the index $\rho$ 
refers to the polarization at the upper end of the gluon line
and $\sigma$ to the lower end. 
$\beta_l$ corresponds to the Sudakov decomposition
$l=\beta_l p + \alpha_l q' + l_t$ where $\alpha_l$ is fixed using the fact that
the quark at the bottom is on-shell ($\alpha_l \simeq l^2_t/s$).
$\beta_l$ itself is given through the on-shell condition of the
intermediate $s$-channel gluon $(l+x_{\fP}p-k)^2=0$ and the final state
gluon $(x_{\fP}p-k)^2=0$:
\beqn\label{app3}
\beta_l&=&\frac{l_t^2-2 l_t\cdot k_t}{\alpha_k s}\\
\alpha_k&=&\frac{k_t^2}{(x_{\fP}-\beta_k)s} \nonumber
\eeqn
($s=2p\cdot q'$).
Here the Sudakov representation of $k$ enters with $\beta_k$ as free
variable denoting the momentum fraction of the upper $t$-channel gluon
with respect to the momentum $p$. Later on it will be substituted by
$z$ ($z=\beta_k/x_{\fP}$) which then denotes the momentum fraction of
the $t$-channel gluon with respect to the Pomeron momentum.  
The contraction of $q'^\sigma$ with the lower quark-gluon vertex gives
roughly $q' \cdot p$ which cancels the same factor in the denominator
of Eq.~(\ref{app2}). The remaining factor $1/(\beta_l+x_{\fP}p)$ in front of the vector
$l_t^\rho$ is large provided that $x_{\fP}$ is small.
The other components of the polarization tensor $d^{\rho \sigma}$ 
are negligible. All these properties are crucial in proving 
the $k_t$-factorization theorem. For the upper $t$-channel gluon the situation
is different. In this case the corresponding tensor reads:
\beqn\label{app4}
d^{\rho\sigma}(k)&=&g^{\rho\sigma}
-\frac{k^\rho q'^\sigma + q'^\rho k^\sigma}{k\cdot q'} \\
&=&g_t^{\rho\sigma}-\frac{k_t^\rho q'^\sigma + q'^\rho k_t^\sigma
+ 2\alpha_k q'^\rho q'\sigma}{\beta_k p\cdot q'} \nonumber
\eeqn
Due to the fact that the contraction of $q'^\sigma$ downwards gives a
factor $x_{\fP} p \cdot q'$ which is not much larger than $\beta_k$,
but of the same order, the term $\beta_k$ in the denominator of
Eq.~(\ref{app4}) is no longer enhanced as in Eq.~(\ref{app2}). However,
a simplification is still possible,
if one restricts oneself to the calculation of leading twist terms and
keeps only the leading logs in $Q^2$. Then, the transverse momenta of the
quarks at the top of the diagram in Fig.~\ref{app1} 
and the gluon below are strongly
ordered and all contributions with an extra inverse power of the large quark
transverse momentum are suppressed. This allows to
set the transverse momentum $k_t$ along any of the quark lines to
zero. Moreover, the projection of $q'^\rho$ with one of the upper
quark-gluon vertices cancels or is sub-leading, and Eq.~(\ref{app4})
may be reduced to: 
\be\label{app5}
d^{\rho\sigma}(k) \;\simeq \;
g_t^{\rho\sigma}-\frac{k_t^\rho q'^\sigma}{\beta_k p\cdot q'}\;\;. 
\ee
This kind of technique is well known and has been applied in deriving
the conventional Altarelli-Parisi splitting function. Therefore it is not
surprising that 
the production of the $q\bar{q}$-system is basically described by the 
AP-splitting function associated with the splitting of a gluon into two quarks
accompanied by a logarithm in $Q^2/k_t^2$. Certainly, this approach
is only valid for the transverse part of the cross section. The
longitudinal part gives a next-to-leading log($Q^2$) contribution
which is not considered here. The coupling of the second gluon to the
$q\bar{q}$-system does not affect the dynamics within this system, but
feels only the total color charge which is the same charge as carried
by the first gluon.

To summarize, the leading twist approach allows to factorize off the
$q\bar{q}$-system analogously to the conventional leading order DGLAP-scheme
whereas in the lower part the $k_t$-factorization theorem is
applicable. All together, a local vertex may be extracted describing
the transition between the lower Pomeron exchange and the upper QCD-radiation.
It is useful to rewrite Eq.~(\ref{app5}) in terms of transverse
polarization vectors $\epsilon_t$ defined as 
\beqn\label{app6}
g_t^{\rho \sigma}&=&-\,\sum_{pol}\;\epsilon_t^\rho \epsilon_t^\sigma \\
(\epsilon_t)_i \cdot (\epsilon_t)_j&=&-\delta_{ij}\;\;.\nonumber
\eeqn
The sum has to be taken over the two helicity or polarization 
configurations in the transverse plane. $d^{\rho \sigma}$ then reads:
\be\label{app7}
d^{\rho \sigma}\;=\;\sum_{pol} \;\epsilon^\rho (k)\;\epsilon^\sigma (k)
\ee
with
\be\label{app8}
\epsilon^\sigma (k)\;=\;\epsilon_t^\sigma\,-\,\frac{k_t \cdot
\epsilon_t \,q'^\sigma}{\beta_k p\cdot q'}\;\;.
\ee

Having in mind the previous discussion one can now start the
calculation of the diagrams in Fig.~\ref{fig_app1}. The effective triple gluon
vertex to the left of the first diagram 
gives the following contribution: 
\beqn\label{app9}
&&2\;\; p\cdot \epsilon(k)\;\; l_t\cdot \epsilon(l+x_{\fP}p-k)\nonumber\\
&-&\frac{\beta_l \alpha_k s}{\beta_l+x_{\fP}}\;\; \epsilon_t\cdot 
\epsilon(l+x_{\fP}p-k)\\
&-&2\;\; p\cdot \epsilon(l+x_{\fP}p-k)\;\; l_t\cdot \epsilon_t \nonumber\\
&-&2\;\; p\cdot \epsilon(l+x_{\fP}p-k)\;\; k_t\cdot \epsilon_t\;
\frac{l_t^2}{\beta_k \alpha_k s} \nonumber
\eeqn
The first three terms of Eq.~(\ref{app9}) result from the ordinary
three gluon coupling whereas the last is the sum of the two
Bremsstrahlungs gluons as illustrated in the first row of Fig.~\ref{fig_app4}. 
The momentum
structure of these contributions is the same except the
overall sign which is opposite. It is obvious that 
the two color tensors add up to the same tensor the 
ordinary three gluon coupling has. The overall color factor will be
evaluated later. Here, only the correspondence between different
color tensors is of interest but not the whole tensor itself.
The right effective vertex in the first diagram of Fig.~\ref{fig_app1} is different as it
contains two $s$-channel gluons. Since these gluons are on-shell, the Ward
identity $l^\rho A_\rho=0$, where $A^\rho$ is the triple gluon coupling
contracted with the gluon polarization vectors, may be used to change 
the $t$-channel polarization vector from $l_t^\rho/\beta_l$ to $p^\rho$. 
The resulting expression is:
\beqn\label{app10}
&-&2\;\; l_t\cdot \epsilon(x_{\fP}p-k)\;\; p\cdot \epsilon(l+x_{\fP}p-k)\nonumber\\
&+&2\;\; l_t\cdot \epsilon(l+x_{\fP}p-k)\;\; p\cdot \epsilon(x_{\fP}p-k) \nonumber\\
&-&\alpha_k s\; \epsilon(l+x_{\fP}p-k)\cdot \epsilon(x_{\fP}p-k)\\
&+&2\;\; p\cdot \epsilon(l+x_{\fP}p-k)\;\; p\cdot \epsilon(x_{\fP}p-k)\;
\frac{l_t^2}{\alpha_k s} \nonumber\;\;.
\eeqn
Both pieces Eqs.~(\ref{app9}) and (\ref{app10}) have to be combined and
the sum over the transverse polarizations of the intermediate $s$-channel gluon
has to be performed. The following equation will be used:
\be\label{app11}
\sum_{pol}\;\epsilon_t^\mu (l+x_{\fP}p-k)\, \epsilon_t^\nu (l+x_{\fP}p-k) \;=\;
-\;g_t^{\mu \nu}\;\;,
\ee
and products like $p \cdot \epsilon (k)$ will be reduced to 
$-2\, k_t\cdot \epsilon_t/\beta_k$. Furthermore, the propagator
$1/k^2=x_{\fP}/(\alpha_k s)=(1-z)/k_t^2$ is introduced and $\beta_l$ is
expressed through Eq.~(\ref{app3}) as well as the variable $\beta_k$
is substituted by $z$ ($\beta_k = x_{\fP} z$):
\beqn\label{app12}
-\,\frac{2}{x_{\fP}}\;\frac{k_t\cdot \epsilon_t}{z k^2}&&\left\{
-\,2 \frac{l_t\cdot
(l_t-k_t)}{(l_t-k_t)^2}\;\left[\frac{l_t^2}{k_t^2} k_t\cdot
\epsilon(x_{\fP}p-k)-l_t\cdot \epsilon(x_{\fP}p-k)\right] \right.\nonumber\\
&&\left. +\,2\,l_t^2\frac{k_t\cdot \epsilon(x_{\fP}p-k)}{k_t^2}\,-\,l_t\cdot
\epsilon(x_{\fP}p-k)\right\}  \\
-\,\frac{1}{x_{\fP}}\left(1-\frac{k^2}{k^2+l_t^2-2l_t \cdot k_t}\right)&&
\left\{-2\frac{(l_t-k_t)\cdot
\epsilon_t}{(l_t-k_t)^2}\left[\frac{l_t^2}{k_t^2} k_t\cdot 
\epsilon(x_{\fP}p-k)-l_t\cdot \epsilon(x_{\fP}p-k)\right]
\right.\nonumber\\ &&\left.+\,2l_t\cdot \epsilon_t \frac{k_t\cdot
\epsilon(x_{\fP}p-k)}{k_t^2}\,-\,\epsilon_t \cdot \epsilon(x_{\fP}p-k) \right\} 
\nonumber \\
+\,\frac{2}{x_{\fP}}\,\left[l_t \cdot \epsilon_t \,+\,
\frac{1-z}{z}\frac{l_t^2}{k_t^2}\,k_t \cdot \epsilon_t \right] 
&&\left\{-\frac{2}{(l_t-k_t)^2}\left[\frac{l_t^2}{k_t^2} k_t\cdot 
\epsilon(x_{\fP}p-k)-l_t\cdot \epsilon(x_{\fP}p-k)\right] \right.
\nonumber\\ 
&&\left.+\,2\,\frac{l_t\cdot (l_t-k_t)}{(l_t-k_t)^2}
\frac{k_t \cdot \epsilon(x_{\fP}p-k)}
{k_t^2}\,-\,\frac{(l_t-k_t)\cdot \epsilon(x_{\fP}p-k)}{(l_t-k_t)^2}\right\}
\nonumber\eeqn

The next contribution has to be taken from the second diagram in
Fig.~\ref{fig_app1}. In this case the situation is slightly simpler
compared to the first diagram,
since only one effective triple gluon vertex appears. Moreover, the
upper $t$-channel gluon is attached to a quark line where the incoming
and the outgoing quarks are on-shell with the consequence that the
momentum of this gluon is purely transverse up to corrections 
proportional to the squared ratio of the gluon transverse momentum 
and the quark transverse momentum. This type of correction is
sub-leading due to the strong ordering assumption. The polarization
tensor simplifies in the following way:
\be\label{app13}
d^{\rho \sigma}(l+x_{\fP}p-k)\;=\;\frac{p^\rho q'^\sigma}{p \cdot q'}\;\;.
\ee 
The upper polarization vector was changed from
$-\frac{(l_t-k_t)^\rho}{\beta_l+x_{\fP}-\beta_k}$ to $p^\rho$ making use
of the fact that the two quarks to the left and to the right are
on-shell. In contrast to the first diagram in Figure \ref{fig_app1} the tensor $g_t^{\rho
\sigma}$ along the $t$-channel line gives only a sub-leading contributions due
to the smallness of the longitudinal momentum. 
The special kinematic situation in the second diagram allows one to
apply the eikonal 
approximation to the right quark-gluon vertex. The subsequent contraction
with $p^\rho$ gives a factor which is cancelled by the residue of the
$\delta$-function corresponding to the intermediate quark, and the remaining
factor is simply $-1$. The softness of the upper right $t$-channel gluon has no
further dynamical effect except that the color charge of both
quarks add up to the total color charge of the left $t$-channel gluon.
Consequently, the color factor is identical to that of the first
diagram in Figure \ref{fig_app1}.
After all, one finds for this diagram:
\beqn\label{app14}
&-&\frac{2}{\beta_k}\;\; l_t\cdot \epsilon_t \;\; 
l_t\cdot \epsilon(x_{\fP}p-k)\\
&+&\frac{2}{\beta_k}\;\; l_t\cdot \epsilon_t \;\;
\frac{l_t^2}{\alpha_k s}\;\; p\cdot \epsilon(x_{\fP}p-k) \nonumber
\eeqn
Inserting the propagator $1/(l_t-k_t)^2$ and substituting $\beta_k$ as
well as $\alpha_k$ one finally comes to:
\be\label{app15}
\frac{2}{x_{\fP}}\;\; \frac{1}{z}\;\; \frac{l_t\cdot\epsilon_t}{(l_t-k_t)^2}\;\; 
\left[\frac{l_t^2}{k_t^2}\;k_t\cdot \epsilon(x_{\fP}p-k)\;-\;
l_t\cdot \epsilon(x_{\fP}p-k)\right]
\ee

In the following step the two expressions (\ref{app12}) and
(\ref{app15}) will be added and the result 
integrated over the azimuth angle between $l_t$ and $k_t$. 
A lot of cancellations occur and the final expression is
rather short:
\beqn\label{app16}
-\;\frac{1}{2x_{\fP}}\;\frac{1}{z(1-z)}&&\left\{z^2\,+\,(1-z)^2\,
+\,\frac{l_t^2}{k^2} \right.\\ 
&&\left.-\,\frac{\left[(1-2z)k^2-l_t^2\right]^2+2z(1-z)k^4}
{k^2\sqrt{(k^2+l_t^2)^2-4(1-z)l_t^2k^2}} \right\}\;\;\epsilon_t \cdot
\epsilon_t(x_{\fP}p-k) \nonumber\;\;.
\eeqn
Recalling the fact that only the amplitude has been considered, the
calculation of the cross section requires to take the square of
Expr.~(\ref{app16}). In doing so one has to sum over the final state
polarizations which leads to a contraction of
the vector $\epsilon_t$ with its conjugate.
In the end the transverse part of the $\gamma$-matrices in the lower edges of 
the quark-box  are contacted as well (see Fig.~\ref{fig_app1}).

Moving on to the final diagram (Fig.~\ref{fig_app1}) one encounters a similar
situation as in the case of the second diagram of the same figure. 
The right $t$-channel
gluon is soft in the sense 
that its momentum is small compared to the quark momenta. It has no
dynamical effect except that the color charge adds up as before, so
that the final color factor is identical to that in the first two diagrams
of Fig.~\ref{fig_app1}. What remains is the calculation of the left
effective triple gluon vertex. This has to be performed in a similar
way as in the case of the left vertex in the second diagram:
\beqn\label{app17}
&&2\;\; p\cdot \epsilon(l+k)\;\; l_t\cdot \epsilon(x_{\fP}p-k)\nonumber\\
&+&\frac{l_t^2+2l_t\cdot k_t}{\beta_l+x_{\fP}}\;\; \epsilon(l+k)\cdot 
\epsilon(x_{\fP}p-k)\\
&-&2\;\; p\cdot \epsilon(x_{\fP}p-k)\;\; l_t\cdot \epsilon(l+k) \nonumber\\
&-&2\;\; p\cdot \epsilon(x_{\fP}p-k)\;\; (l_t+k_t)\cdot \epsilon(l+k)\;
\frac{l_t^2}{(\beta_l+\beta_k) \alpha_k s} \nonumber
\eeqn
The last term in Eq.~(\ref{app17}) summarizes the contribution of the
Bremsstrahlungs gluons associated with the effective triple gluon
coupling. As was argued before the longitudinal momentum of the right
soft $t$-channel gluon is negligible and $\beta_l$ equals zero.
The momentum of the upper left $t$-channel gluon does not reduce to its
transverse component, but includes the non-negligible longitudinal
fraction $z$ of the Pomeron momentum. Therefore, the propagator
$1/(l+k)^2$ transforms into $1/(zk^2+(l_t+k_t)^2)$. Introducing this
propagator into Eq.~(\ref{app17}) and substituting $\beta_k=zx_{\fP}$ as well
as $\alpha_k s=k^2/x_{\fP}$ one finds:
\beqn\label{app18}
&-&\frac{2}{x_{\fP}}\;\frac{1}{z}\; \frac{(l_t+k_t)\cdot \epsilon(l+k) 
\; l_t\cdot \epsilon(x_{\fP}p-k)}{zk^2+(l_t+k_t)^2}\nonumber\\
&-&\frac{2}{x_{\fP}}\;\frac{1}{z}\; \frac{l_t^2}{k_t^2}\;\; 
\frac{(l_t+k_t)\cdot \epsilon(l+k) 
\; k_t\cdot \epsilon(x_{\fP}p-k)}{zk^2+(l_t+k_t)^2}\nonumber\\
&-&\frac{2}{x_{\fP}}\;\frac{1}{1-z}\; \frac{l_t\cdot \epsilon(l+k) 
\; k_t\cdot \epsilon(x_{\fP}p-k)}{zk^2+(l_t+k_t)^2} \\
&+&\frac{1}{x_{\fP}}\;\frac{(l_t+k_t)^2-k_t^2}{zk^2+(l_t+k_t)^2}\;\; 
\epsilon(l+k)\cdot \epsilon(x_{\fP}p-k)\nonumber
\eeqn
Once more one has to integrate over the azimuth angle between $l_t$
and $k_t$ with the remarkable outcome that the resulting expression is
identical to Eq.~(\ref{app16}):
\beqn\label{app19}
-\;\frac{1}{2x_{\fP}}\;\frac{1}{z(1-z)}&&\left\{z^2\,+\,(1-z)^2\,
+\,\frac{l_t^2}{k^2} \right.\\ 
&&\left.-\,\frac{\left[(1-2z)k^2-l_t^2\right]^2+2z(1-z)k^4}
{k^2\sqrt{(k^2+l_t^2)^2-4(1-z)l_t^2k^2}} \right\}\;\;\epsilon_t \cdot
\epsilon_t(x_{\fP}p-k) \nonumber\;\;.
\eeqn
In other words, the sum of the first two diagrams in Fig.~\ref{fig_app1} 
is identical
to the third diagram bearing in mind that the light
cone gauge with the condition $q'\cdot A=0$ was used. One should
remind that the amplitude was calculated in the high energy asymptotic
region where the real parts of the $s$-channel and $u$-channel contributions
cancel due to the even signature of the color singlet exchange. (The
$u$-channel contribution corresponds to the crossing of the two lower
$t$-channel gluons in Fig.~\ref{fig_app1}.). Hence, the imaginary part gives
the leading part and was calculated taking the $s$-channel
discontinuity, i.e. cutting the diagrams. However, the cut diagram gives
twice the imaginary part and one has to divide the final result by 2.

The structure in Eq.~(\ref{app19}) has been used in Eq.~(\ref{diff3}). 
The wave function in Eq.~(\ref{gluondipole}) cannot be extracted directly from 
the diagrams discussed here, but was constructed such  
that it reproduces the same results.

%%%%%%%%%%%%%%%%%%%%%%%%%%%%%%%%%%%%%%%%%%%%%%%%%%%%%%%%%%%%%%%%%%

%%%%%%%%%%%%%%%%%%%%%%%%%%%%%%%%%%%%%%%%%%%%%%%%%%%%%%%%%%%%%%%%%%%%
\newpage
%\input fig_rest
%%%%%%%%%%%%%%%%%%%%%%%%%%%%%%%%

\newpage  

\newpage  

\newpage   

\newpage   

\newpage   

\newpage   
%%%%%%%%%%%%%%%%%%%%%%%%%%%%%%%%%%%%% 
\end{document}